\documentclass[apj]{emulateapj}
\usepackage{url}
\usepackage{lscape}
\usepackage{amsmath}
\usepackage[]{hyperref}
\hypersetup{colorlinks=true,citecolor=blue,filecolor=black,runcolor=blue,linkcolor=blue,urlcolor=blue}
\shortauthors{MORISHITA, ICHIKAWA \& KAJISAWA.}
\begin{document}

\title{The Evolution of Galaxy Size and Morphology at $z\sim$ 0.5--3.0 in the GOODS-N region with $HST$/WFC3 Data} 

\author{Takahiro Morishita, Takashi Ichikawa}
\affil{Astronomical Institute, Tohoku University, Aramaki, Aoba, Sendai 980-8578, Japan}
\email{mtakahiro@astr.tohoku.ac.jp}

\and 

\author{Masaru Kajisawa}
\affil{Research Center for Space and Cosmic Evolution, Ehime University, Bunkyo-cho, Matsuyama 790-8577, Japan}

%%%%%%%%%%%%%%%%%%%%%%%%%%%%%%%%%%%%%%%%%%%%%%%

\begin{abstract}
We analyze the recent released $HST$/WFC3 IR images in the GOODS-N region to study the formation and evolution of Quiescent galaxies (QGs). 
After examining the reliability with artificial galaxies, we obtain the morphological parameters with S\'ersic profile of 299 QGs and 1,083 star-forming galaxies (SFGs) at $z\sim0.5$--3.0, finding the evolution of $r_\mathrm{e}$ and $n$ of massive ($M_*\geq10^{10.5}$~M$_\sun$) QGs while weaker evolution of SFGs and less massive ($M_*<10^{10.5}$~M$_\sun$) QGs.
The regression of the size evolution of massive QGs follows $r_\mathrm{e}\propto (1+z)^{-\alpha_{r_\mathrm{e}}}$ with $\alpha_{r_\mathrm{e}}=1.06\pm0.19$ (a factor of $\sim2.2$ increase from $z\sim2.5$ to $\sim0.5$), which is consistent with the general picture of the significant size growth.
For the further understanding of the evolution scenario, we study the evolution of S\'ersic index, $n$, and find that of massive QGs to significantly evolve as $n\propto(1+z)^{-\alpha_n}$ with $\alpha_n=0.74\pm0.23$ ($n\sim1$ at $z\sim2.5$ to $n\sim4$ at $z\sim0.5$), while those of the other populations are unchanged ($n\sim1$) over the redshift range.
The results in the present study are consistent with both of observation and numerical simulations, where gas-poor minor merger is believed to be the main evolution scenario.
By taking account of the connection with less massive QGs and SFGs, we discuss the formation and evolution of the massive QGs over{\it ``Cosmic High Noon"}, or the peak of star-formation in the universe.

\end{abstract}
\keywords{galaxies: evolution - galaxies: high-redshift - galaxies: elliptical and lenticular, cD - galaxies: structure}

%%%%%%%%%%%%%%%%%%%%%%%%%%%%%%%%%%%%%%%%%%%%%%%

\section{Introduction}
The study of the high-redshift (high-$z$) early-type galaxies (ETGs) provides us clues to understanding the formation and evolution of massive galaxies in the local universe. 
Their star-formation activity peaked during the cosmological epoch at $1 < z < 3$ (e.g., Dickinson et al.~\citeyear{dickinson03}; Heavens et al.~\citeyear{heavens04}; Papovich et al.~\citeyear{papovich06}; Hopkins \& Beacom~\citeyear{hopkins06}) and galaxy morphologies have changed dramatically (Kajisawa \& Yamada~\citeyear{kajisawa06}). 
For galaxy sizes, many studies have corroborated that massive galaxies at high-$z$ were much smaller than local galaxies with comparable mass (Daddi et al.~\citeyear{daddi05}; Trujillo et al.~\citeyear{trujillo06},~\citeyear{trujillo07}; Cimatti et al.~\citeyear{cimatti08}; van~Dokkum et al.~\citeyear{vandokkum08}; Akiyama et al.~\citeyear{akiyama08}; Franx et al.~\citeyear{franx08}; Szomoru et al.~\citeyear{szomoru10},~\citeyear{szomoru12}; van~der~Wel et al.~\citeyear{vanderwel11}; Barro et al.~\citeyear{barro13}). 
At a fixed stellar mass, ETGs are claimed to have been significantly compact at high-$z$ and have evolved with rapid increase of their effective radius by a factor of $\sim$ 4 or even larger from $z\sim 2$ (Buitrago et al.~\citeyear{buitrago08}; Carrasco et al.~\citeyear{carrasco10}) and by a factor $\sim$ 2 from $z \sim 1$ (van~der~Wel et al.~\citeyear{vanderwel08}; Trujillo et al.~\citeyear{trujillo11}). 
To reach the size of local ETGs, rapid and violent evolutions by major merger (Hopkins et al.~\citeyear{hopkins09a}) or minor merger (Bezanson et al.~\citeyear{bezanson09}; Naab et al.~\citeyear{naab09}) have been demanded. 

Recent very deep infrared observations of high spatial resolution with the $Hubble\ Space\ Telescope\ (HST)$ have shed light on morphological details and shapes of galaxies at high-$z$. 
Bruce et al.~(\citeyear{bruce12}) studied over 200 massive galaxies at $1 < z < 3$ in the CANDELS-UDS field and found that these galaxies had much smaller size at a given mass than that of local ETGs. 
On the other hand, it has also been argued that the compact galaxies have apparent smaller effective radii because of low signal to noise ratio (S/N) (e.g., Ryan et al.~\citeyear{ryan12}). 
The lack of the consideration for AGN component would also make the radius smaller (Yoshino \& Ichikawa~\citeyear{yoshino08}; Pierce et al.~\citeyear{pierce10}). 
In addition, the best-fit morphological outputs with, for example, {\ttfamily GALFIT} (Peng et al.~\citeyear{peng02}), which is one of the most frequently used fitting codes for galaxy morphology, could be significantly changed with small differences of fitting inputs (e.g., initial guess, point spread function (PSF), weight image) and the image properties (e.g., size of postage stamp, sky background noise).
Although {\ttfamily GALFIT} are frequently used, it sometimes gives inappropriate results, mostly when used without careful considerations to image quality of galaxies and to the contamination by neighboring objects (H\"{a}u{\ss}ler et al.~\citeyear{haussler07}, hereafter H07; Barden et al.~\citeyear{barden12}).

Some previous studies (e.g., Trujillo et al.~\citeyear{trujillo06}; H07; Carollo et al.~\citeyear{carollo13}; Mosleh et al.~\citeyear{mosleh13}) estimated the errors in effective radius, $r_\mathrm{e}$, and S\'ersic index, $n$, using artificial galaxies (AGs), and derived simple relations between the original and output values. 
Szomoru et al.~(\citeyear{szomoru10}) contrived to compensate the faint extended wings of galaxies. 
They estimated the limit of surface brightness and fitted the S\'ersic profile to the galaxy images above the surface brightness limit with {\ttfamily GALFIT}. 
Then, they corrected the result $r_\mathrm{e}$ by calculating the residual counts between the original and model images. 
Based on a careful study of the bias of image quality and PSF profiles, van~der~Wel et al.~(\citeyear{vanderwel12}) presented global structural parameters of more than 100,000 galaxies in the CANDELS survey. 
Bruce et al.~(\citeyear{bruce12}) applied {\ttfamily GALFIT} to three-component fitting (bulge, disk, and central components) for high-$z$ galaxies with a careful attention to the background noise and PSF convolution.

In addition, we should take account of the different analysis for local galaxies when comparing the morphological properties at high-$z$. 
The half-light radii of the SDSS local galaxies used in Shen et al.~(\citeyear{shen03}, hereafter S03) were claimed to be underestimated (Guo et al.~\citeyear{guo09}, hereafter G09; Simard et al.~\citeyear{simard11}). 
The comparison of the size-stellar mass relations between the different definitions of stellar mass would also lead to inappropriate results (Mosleh et al.~\citeyear{mosleh13}). 
The comparison of the structural parameters for high-$z$ galaxies with those in the local universe should be based on the consistent definition and analysis of galaxy data. 

In this paper, we investigate the reliability and limit of {\ttfamily GALFIT} to obtain the morphological properties of high-$z$ galaxies. 
Then, using deep near infrared (NIR) observations with Wide Field Camera 3 (WFC3) instrument installed on $HST$, we apply {\ttfamily GALFIT} to galaxies in the Great Observatories Origins Deep Surveys-North (GOODS-N) region, in which Ichikawa et al.\ (\citeyear{ichikawa12}) (hereafter Ic12) studied the size evolution of galaxies in a non-parametric way with $K_\mathrm{s}$-band ground-based images of MOIRCS Deep Survey (MODS). 
The ground-based images were not reliable enough for the morphological study of galaxies at $z>1$ with {\ttfamily GALFIT} (Konishi et al.~\citeyear{konishi11}).
As such, Ic12 obtained the size-stellar mass relations based on half- and 90 percent light radii.
On the other hand, deep images by WFC3 with much higher spatial resolution will allow us to apply {\ttfamily GALFIT} to high-$z$ galaxies, including compact galaxies, for the morphological study.

An outline of the paper is as follows.
In Section~\ref{sec:sec2}, we describe the samples of massive galaxies in the GOODS-N.
Using the background noise and PSF of WFC3 images, we make AGs with various shape parameters. 
We analyze them with {\ttfamily GALFIT} and compare the results with the original parameters in Section~\ref{sec:sec3}. 
We examine the reliability and the systematic errors of $r_\mathrm{e}$ and $n$ of the AGs obtained with {\ttfamily GALFIT} under some conditions. 
After examining the validity of the results, we apply the fitting method to massive galaxies in the GOODS-N. 
The results are described in Section~\ref{sec:sec4} and Section~\ref{sec:sec5}. 
Finally we discuss our results in comparison with those of previous parametric and non-parametric studies in Section~\ref{sec:sec6}. 
Throughout this paper, we assume $\Omega_m$ = 0.3, $\Omega_\mathrm{\Lambda}$ = 0.7 and $H_0$ = 70 kms$^{-1}$Mpc$^{-1}$. 
We use the AB magnitude system (Oke \& Gunn~\citeyear{oke83}; Fukugita et al.~\citeyear{fukugita96}).

%%%%%%%%%%%%%%%%%%%%%%%%%%%%%%%%%%%%%%%%%%%%%%%%%%%%%

\section{Data} \label{sec:sec2}
We use the NIR data taken with $HST$/WFC3 in the CANDELS survey (Grogin et al.~\citeyear{grogin11}; Koekemoer et al.~\citeyear{koekemoer11}). 
The survey targeted approximately 120 arcmin$^2$ to 10-epoch depth in $J_\mathrm{125}$ and $H_\mathrm{160}$ in the GOODS-N region.  
We use the full data of the observations through the Mikulski Archive for Space Telescopes (MAST). 
$J_\mathrm{125}$ and $H_\mathrm{160}$ correspond to rest-frame $V$-band images at $z\sim$1.0--1.8 and 1.8--3.0, respectively. 
The images are reduced through {\ttfamily PyRAF} package ($DrizzlePac$, Gonzaga et al.~\citeyear{gonzaga11}), where the standard calibrations (i.e. flat, sky background subtraction, distortion correction, cosmic ray rejection) are done. 
The images are drizzled to a pixel size of $0\arcsec\!.06$ using pixel fraction value of 0.8 to be consistent with published GOODS-S images (Koekemoer et al.~\citeyear{koekemoer11}). 
The full width at half-maximums (FWHMs) of the PSF are $\sim 0\arcsec\!.15$ and $0\arcsec\!.18$ for $J_\mathrm{125}$ and $H_\mathrm{160}$ images, respectively, which are estimated by median stacked unsaturated stars. 
The PSF for $H_{160}$ image is consistent with that of the GOODS-S images, though that of $J_\mathrm{125}$ is slightly larger. 
As the PSF profile is one of the most important parameters for the morphological analysis of galaxies, we investigate it carefully in Section~\ref{sec:sec4.2}.

%%%%%%%%
% Graphics %
%%%%%%%%
\begin{figure}
\figurenum{1a}
\begin{center}
\plotone{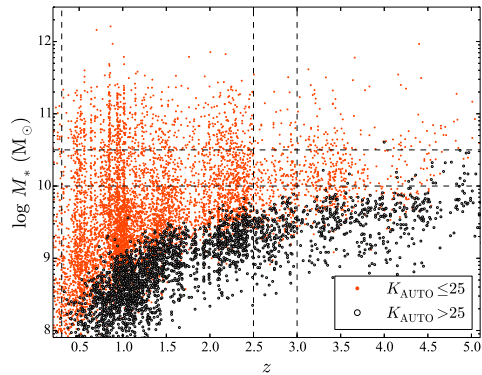}
\caption{
MODS $K_\mathrm{s}$-band selected galaxies with $K_\mathrm{AUTO}\leq25$ (red dots) and $K_\mathrm{AUTO}>25$ (open circles).
The vertical and the horizontal dash lines show the redshift and stellar mass limits for the present study, respectively.
}
\label{fig:fig1a}
\end{center}
\end{figure}

\begin{figure*}
\figurenum{1b}
\begin{center}
\plotone{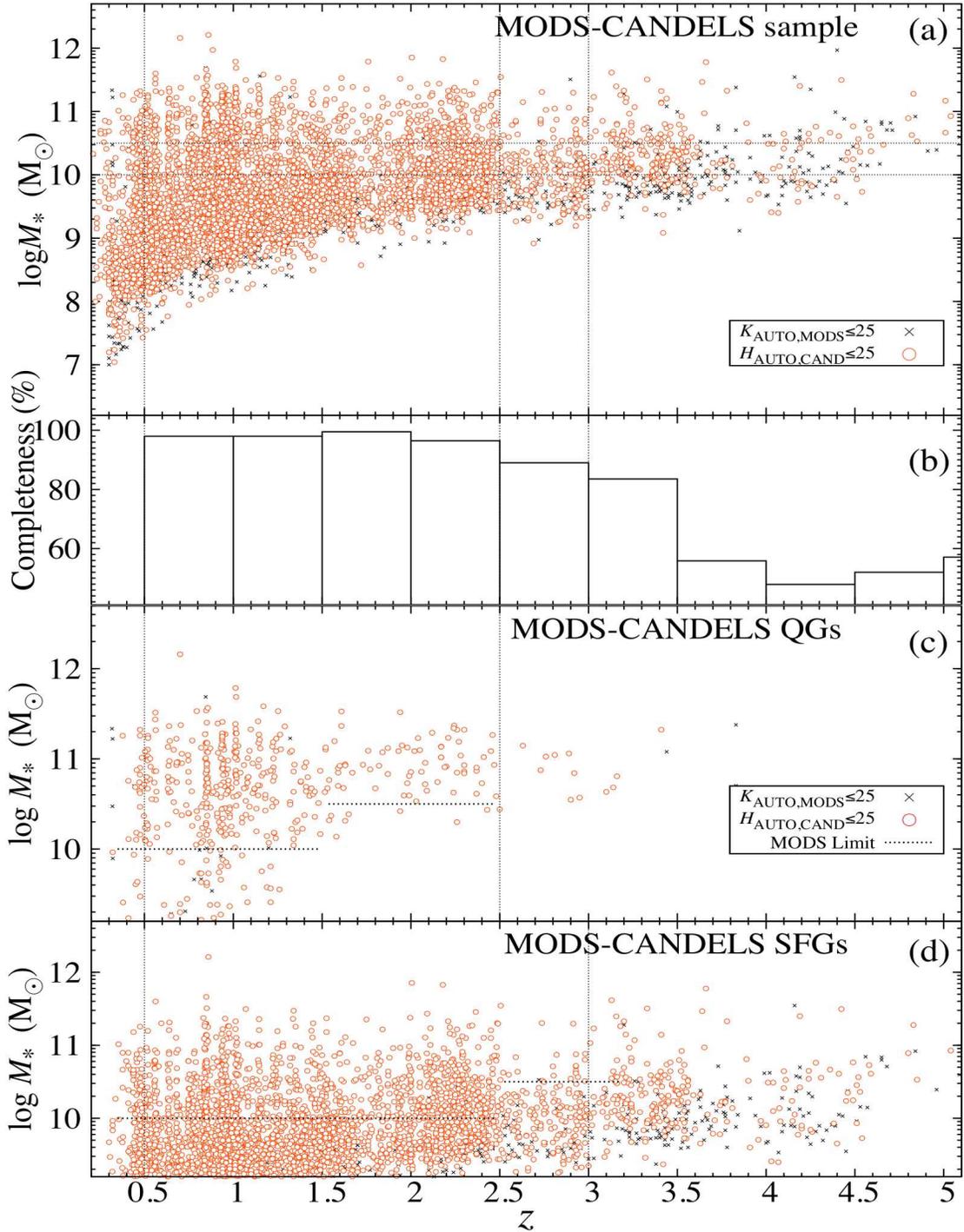}
\caption{
{\bf (a)} Redshift-stellar mass diagram for the present sample with $H_\mathrm{AUTO}\leq25$.
$H_\mathrm{AUTO}$ obtained by {\ttfamily SExtractor} on the WFC3/$H_{160}$ image are depicted as red open circles.
Black crosses represent MODS samples with $K_\mathrm{AUTO}\leq25$, which are not detected with $H_\mathrm{AUTO}\leq25$.
The vertical and the horizontal dash lines show the redshift and stellar mass limits for the present study, respectively.
{\bf (b)} Completeness for the present sample with $M_*\geq10^{10}$~M$_{\sun}$, where the ordinate represents the completeness in percentage of the galaxies with $H_\mathrm{AUTO}\leq25$ to those with $K_\mathrm{AUTO}\leq25$ and $M_*\geq10^{10}$~M$_{\sun}$.
{\bf (c)} Redshift-stellar mass diagrams for QGs with $H_\mathrm{AUTO}\leq25$, where QGs are selected based on rest-frame $UVJ$ color (see $\S5.1$).
The horizontal dash lines show the stellar mass limits for QGs for each redshift bin.
{\bf (d)} Same as (c) but for SFGs.
}
\label{fig:fig.1}
\end{center}
\end{figure*}

We make use of the $K_\mathrm{s}$-band selected catalog of the MODS in the GOODS-N region (Kajisawa et al.~\citeyear{kajisawa09}, hereafter K09; Kajisawa et al.~\citeyear{kajisawa11a}, K11), which is based on the imaging observations in $J$-, $H$- and $K_\mathrm{s}$-bands with MOIRCS (Ichikawa et al.~\citeyear{ichikawa06}; Suzuki et al.~\citeyear{suzuki08}) installed on the Subaru telescope. 
To obtain the stellar mass ($M_*$) of MODS samples, K09 performed SED fitting of multi-band photometry ($UBVizJHK_\mathrm{s}, 3.6{\micron}, 4.5{\micron}$ and $5.8{\micron}$) with population synthesis models. 
They adopted the results with GALAXEV templates (Bruzual \& Charlot~\citeyear{bruzual03}) and Salpeter (\citeyear{salpeter55}) initial mass function (IMF). 
The stellar masses are obtained from the best fit stellar-mass to luminosity ratio in $K_\mathrm{s}$-band and scaled with the $K_\mathrm{s}$-band flux. 
In the catalog, 2,093 of 9,937 galaxies have spectroscopic redshifts. 
The photometric redshifts were calculated for other galaxies from the best fit SED models. 
The derived photometric redshift showed excellent agreement with the spectroscopic redshift $\delta z/(1+z)=-$0.011$\pm$0.078 (see K11 for more details). 
We use the spectroscopic redshifts if available, and the photometric redshifts for others.
\textcolor{black}{
In K11, they estimated the completeness $>85$\% at 5~$\sigma$ detection limit of the image ($K_\mathrm{AUTO}\sim26$~mag for deep region and $\sim25$~mag for wide, where $K_\mathrm{AUTO}$ is {\ttfamily SExtractor} MAG\_AUTO; see Section~\ref{sec:sec3.2} for details) by using Monte-Carlo simulations for the artificial galaxies buried in the MODS $K_\mathrm{s}$-band image.
In the present study, we follow the detection limit of $K_\mathrm{AUTO}\leq25$.
In Fig.~\ref{fig:fig1a}, we see that all galaxies with $M_*\geq10^{10}$~M$_{\odot}$ at $0.5\leq z\leq3.0$ have $K_\mathrm{AUTO}\leq25$, which promises the completeness (of $>85\%$) for the sample in the stellar mass and redshift ranges.
}

From MODS sample, we select quiescent galaxies (QGs) and star-forming galaxies (SFGs) based on rest frame $UVJ$ color (see Section~\ref{sec:sec5.1} for details).
Since the completeness in redshift bins for the two galaxy populations are different, we set the redshift limit: $M_*\geq10^{10}M_\mathrm{\odot}$ for QGs (SFGs) at $0.5\leq z\leq1.5$ ($0.5\leq z\leq2.5$) and $\geq10^{10.5}M_\mathrm{\odot}$ at $1.5\leq z\leq2.5$ ($2.5\leq z\leq3.0$) (see Kajisawa et al.~\citeyear{kajisawa11b} for more detail).
The whole samples of QGs and SFGs in redshift-stellar mass diagrams with the completeness are shown in Fig.~\ref{fig:fig.1}, where we see that over 90\% of the MODS galaxies are included in the present sample (see Section~\ref{sec:sec3.3} about the selection limit).

%%%%%%%%%%%%%%%%%%%%%%%%%%%%%%%%%%%%%%%%%

\section{Morphological Analysis}\label{sec:sec3}
Many studies for galaxy morphologies are based on finding the best-fit 2D surface brightness profile of S\'ersic\ (\citeyear{sersic68}), which is written as
\begin{equation}
I(r)=I_\mathrm{e}\exp\left[-b(n)\left(\left(\frac{r}{r_\mathrm{e}}\right)^{1/n}-1\right)\right],\\
\end{equation}
where $n$ is S\'ersic index, $r_\mathrm{e}$ effective radius and $I_\mathrm{e}$ surface brightness at $r_\mathrm{e}$. 
$b(n)$ is defined as a function of $n$. 
To fit galaxies with S\'ersic profile, we use {\ttfamily GALFIT}, considering the effect of PSF, background noise, and frame size of fitting, for the morphological parameters.
{\ttfamily GALFIT} sometimes returns unpleasant or biased results when used without much care to, for example, sky subtraction, initial guess, and weight images. 
To examine the bias and uncertainties of the results, we prepare AGs and apply {\ttfamily GALFIT} to them. 
Another concern is neighboring galaxies or stars. 
For the galaxies with neighbors, we apply {\ttfamily GALFIT} after masking the neighboring objects, as done by many studies\footnotemark.

\footnotetext{You can see the details about our original script ({\ttfamily SEROGANS EX}) in the following.\\
\href{http://www.astr.tohoku.ac.jp/~mtakahiro/sci/SEROGANS.html}{http://www.astr.tohoku.ac.jp/$\sim$mtakahiro/sci/SEROGANS.html}}

\subsection{Fitting with GALFIT} \label{sec:sec3.1}
{\ttfamily GALFIT} is a 2D fitting code which calculates $\chi^2$ for model galaxies and finds the model with minimum $\chi^2$. 
It is frequently used for the discussion on morphologies.
Trujillo et al.~(\citeyear{trujillo06}) used {\ttfamily GALFIT} for AGs and estimated errors in $r_\mathrm{e}$ and $n$.
Comparing {\ttfamily GALFIT} and {\ttfamily GIM2D} (Simard~\citeyear{simard02}), H07 concluded that the former gave better results for faint galaxies. 
In addition, Ravindranath et al.~(\citeyear{ravindranath06}) and  Cimatti et al.~(\citeyear{cimatti08}) showed that {\ttfamily GALFIT} returned unbiased estimates of $r_\mathrm{e}$ and $n$ for galaxies with S/N $>$ 10 and $r_\mathrm{e}>$ 0\arcsec\!.03 at any redshifts of the sources. 
Szomoru et al.\ (\citeyear{szomoru12}) used masks on outer faint part of galaxies in order to exclude the sky background noise at the extend wings. 

As described above, {\ttfamily GALFIT} is widely used for morphological analyses of galaxies. 
In the following sections, we examine the reliability in a similar context to the methods of previous studies (e.g., H07).
We use the newest version of {\ttfamily GALFIT} 3.0.5, which is improved in creating weight images. 
The revision makes results more reliable for faint objects (C. Peng, private communication). 

\subsection{The Initial Guess for Input Parameters} \label{sec:sec3.2}
To perform morphological fitting with {\ttfamily GALFIT}, we should provide an appropriate set of initial morphological parameters, which leads to more reliable parameters and saves CPU cost. 
As done in previous studies, we use {\ttfamily SExtractor} version 2.5.0 (Bertin \& Arnouts~\citeyear{bertin96}) to estimate galaxy properties of position, magnitude, radius, axis ratio, and position angle of the target. 
Total magnitude, MAG\_AUTO ($m_\mathrm{AUTO}$), and half-light radius, FLUX\_RADIUS 50 ($r_{50}$), which encircles half the light emitted from galaxies, are the initial guess of  magnitude and effective radius, respectively. 
No constraints are imposed on the parameters during fitting.

One problem still remains; how can we set the initial $n$? 
To avoid the problem, we use the initial $n$ from 0.5 to 8 by step of 0.5 and iterate {\ttfamily GALFIT} 16 times for each galaxy to derive more reliable parameter sets, as employed by Bruce et al.~(\citeyear{bruce12}). 
After getting results, we compare (at most 16 of) "best-fit" parameter sets based on their reduced chi square, $\chi^2/\nu$, where $\nu$ is the number of degrees of freedom for fitting, though we discard the results with unrealistic parameters (see Section~\ref{sec:sec3.3} for the details).
If a galaxy has different parameter sets with similar $\chi^2/\nu$, we adopt the set which has the model magnitude nearest to $m_\mathrm{AUTO}$, because the selection based on $m_\mathrm{AUTO}$, which is independent of the parametric profile, gives more reliable parameter sets.

\subsection{Fitting Test for AGs in the WFC3 Images} \label{sec:sec3.3}
As we mentioned above, we should scrutinize the reliability {\ttfamily GALFIT} with AGs before applying to the real galaxies in the GOODS-N region. 
AGs are prepared by {\ttfamily GALFIT} with random sets of $r_\mathrm{e}$, $n$, total magnitude ($m$), $b/a$ and position angle ($PA$) (see Table~1). 
$r_\mathrm{e}$ is circularized as $r_\mathrm{e} = a_\mathrm{e} \sqrt{b/a}$, where $a_e$ is the effective radius along the semi-major axis derived by {\ttfamily GALFIT}. 
The images are convolved with a Moffat-profile PSF of FWHM $\sim0\arcsec\!.16$ and $\beta$ = 2.5. 
{\ttfamily GALFIT} precisely returns the original parameters if images have no noise.
We also make AGs using {\ttfamily IRAF} packages  {\it gallist} and {\it mkobjects} with the same parameter sets and PSF. 
{\ttfamily GALFIT} works well again on them if there is no noise. 
It is noted, however, that AGs made by {\ttfamily IRAF} are different from those by {\ttfamily GALFIT}, especially in the center part when they are convolved with PSF.
The difference could originate from the systematic inconsistency of the convolution method, as noted in GALFIT Q\&A\footnotemark, while templates of {\ttfamily GALFIT} convolved with PSF are in good agreement with those of {\ttfamily GIM2D}.
\footnotetext{\href{http://users.obs.carnegiescience.edu/peng/work/galfit/TFAQ.html}{http://users.obs.carnegiescience.edu/peng/work/galfit/TFAQ.html}}
Therefore, we use AGs made by {\ttfamily GALFIT} throughout this paper. 

\begin{figure}
\figurenum{2a}
\plotone{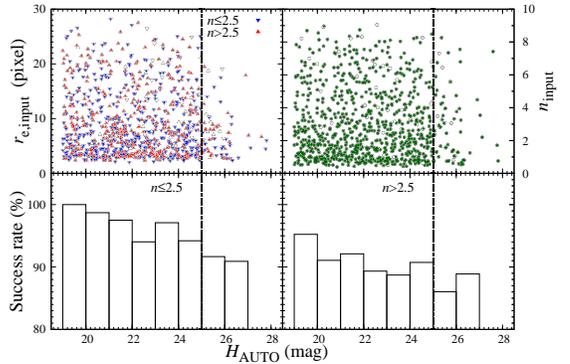}
\caption{
{\it Top} : Initial parameters for artificial galaxies (AGs). 
The AGs fitted successfully with {\ttfamily GALFIT} are depicted with blue ($n\leq 2.5$) and red ($n >$2.5)  triangles for $r_\mathrm{e}$ (left) and green filled circles for $n$ (right) as a function of $H_\mathrm{AUTO}$.
The failed results are shown with open symbols. 
{\it Bottom} : Success rate of {\ttfamily GALFIT} for AGs with $n_\mathrm{input}\leq2.5$ (left) and $n_\mathrm{input}>2.5$ (right).
The definition for success rate is described in the text.
The vertical lines at $H_\mathrm{AUTO}=25$ represent the limit for the present study.
The input parameters are summarized in Table~1.
}
\label{fig:fig.2a}
\end{figure}

\begin{figure*}
\figurenum{2b}
\plotone{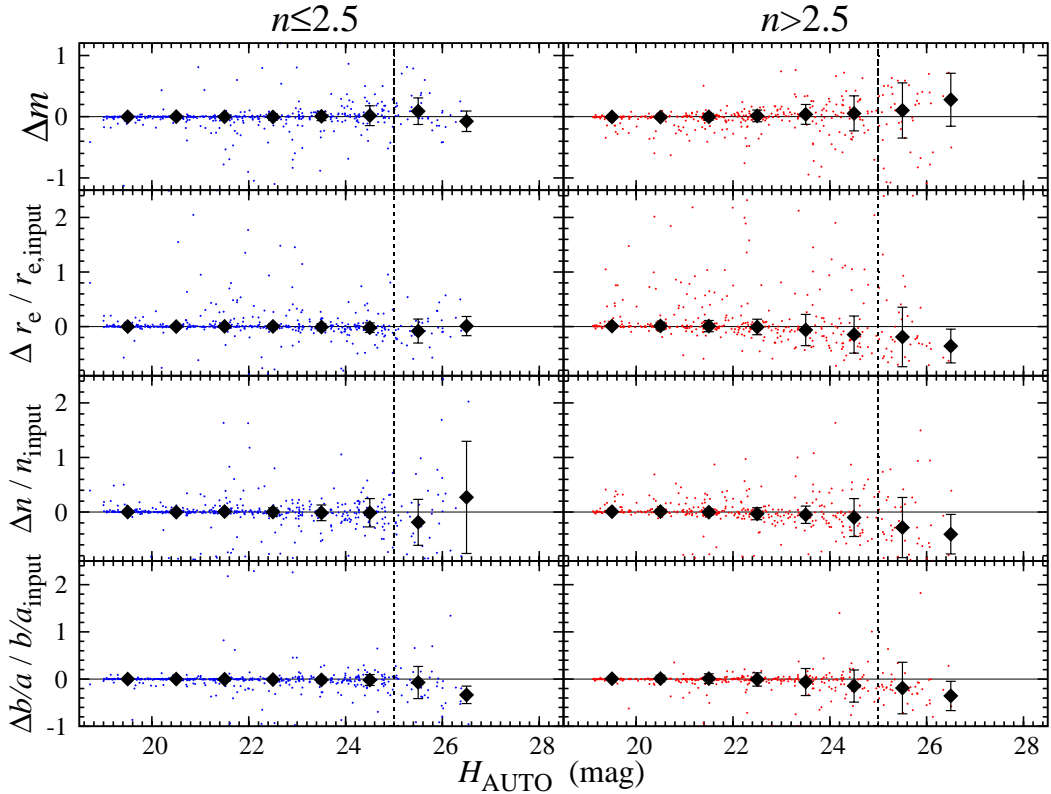}
\caption{
Results of {\ttfamily GALFIT} for artificial galaxies (AGs) buried in the WFC3 $H_{160}$ images. 
The abscissa is the $H$-band MAG\_{AUTO} ($H_\mathrm{AUTO}$) derived by {\ttfamily SExtractor}. 
The ordinate is the difference of the input and output parameters defined as $\Delta m = m_\mathrm{output}-m_\mathrm{input}$, $\Delta r_\mathrm{e}/r_\mathrm{e,input}=(r_\mathrm{e,output}-r_\mathrm{e,input})/r_\mathrm{e,input}$, $\Delta n/n_\mathrm{input}=(n_\mathrm{output}-n_\mathrm{input})/n_\mathrm{input}$ and $\Delta (b/a)/(b/a)_\mathrm{input}=(b/a_\mathrm{output}-{b/a}_\mathrm{input})/(b/a)_\mathrm{input}$.
The median values for each magnitude bin are depicted with filled circles with their median absolute dispersions (MADs). 
The result values are summarized in Table~2.
}
\label{fig:fig.2b}
\end{figure*}

After making PSF-convolved AGs with poisson noise ({\ttfamily IRAF}/$mknoise$), we bury them in the WFC3 images at random positions. 
Although WFC3 images are sky-subtracted through the drizzle tasks, we repeat the sky subtraction for each postage stamp to remove the local sky anomaly evaluated with {\ttfamily IRAF}/$imstat$. 
In order to estimate the sky background and to create the sigma image, {\ttfamily GALFIT} demands larger postage stamps for larger galaxies. 
\textcolor{black}{
Therefore, for each galaxy we cut a square image from the mosaic images with a side of $r_\mathrm{fit}$ in pixel,
\begin{equation}
r_\mathrm{fit}=2(3a Kron+20),\\
\end{equation}
where $a$ is the {\ttfamily SExtractor} output A\_IMAGE and $Kron$ is KRON\_RADIUS.
$r_\mathrm{fit}$ is large enough for applying {\ttfamily GALFIT} to the present samples ($r_\mathrm{fit}\leq500$ pixels).
}
The contamination by the neighbors easily affects the result of {\ttfamily GALFIT} (H07), and therefore it should be removed. 
The neighboring objects around the target galaxies are masked out with SEGMENTATION\_IMAGE obtained by {\ttfamily SExtractor}. 
The results that {\ttfamily GALFIT} does not converge or with inconsistent magnitudes between {\ttfamily GALFIT} and {\ttfamily SExtractor}, $|\Delta m| = |m_\mathrm{GALFIT}-m_\mathrm{AUTO}| > 1.0$ are discarded.
If neighboring objects are detected within 10 pixels from the target galaxy, we also discard the target from final results to avoid possible biases (H\"{a}u{\ss}ler et al.~\citeyear{haussler13}). 
We refer to the rate of galaxies for which {\ttfamily GALFIT} converges and fulfills the above criteria as success rate.
It is noted that masking images sometimes cause systematic errors in the result. 
The errors are, however, much less significant than those due to the disturbance by neighboring objects.
On top of it, masking the neighbors enables {\ttfamily GALFIT} to increase the successful rate. 
The results of {\ttfamily GALFIT} for PSF-convolved AGs with noise are shown in Figs.~\ref{fig:fig.2a} and \ref{fig:fig.2b}.
In Fig.~\ref{fig:fig.2a}, we show the input parameters for successful and failed results.
The success rate tends to be smaller for faint galaxies, mainly because of the disturbance by neighboring objects.
Since {\ttfamily GALFIT} returns unbiased results for AGs with $H_\mathrm{AUTO}\leq25$ as found in Fig.~\ref{fig:fig.2b}, we set the limit magnitude $H_\mathrm{AUTO}=25$ for the analysis of real galaxies in the GOODS-N region.
This limit covers over 90\% for our sample at redshift bins of $z\leq3.0$ (see Fig.~\ref{fig:fig.1}).
If AGs are buried in a clean postage stamp without any bright neighbors, not at random place, in the GOODS-N images, {\ttfamily GALFIT} returns a higher success rate and less biased results even at fainter magnitudes.

It is noted that we use the sigma image defined by {\ttfamily GALFIT} to estimate $\chi^2$ values, because the sigma image created by the drizzle task is questionable for the present analysis, as mentioned by Gonzaga et al.~(\citeyear{gonzaga11}).
Using the sigma images of {\ttfamily GALFIT} also keeps the consistency between the simulation of AGs and the analysis of real galaxies. 
As described in {\ttfamily GALFIT} manual, we use the image in a unit of [count] rather than [count/second], to properly create sigma images. 
The usage of images in [count/second] would give biased results (see GALFIT Q\&A).

%%%%%%%%
% Graphics %
%%%%%%%%

\begin{deluxetable}{ccccc}
\tablenum{1}
\tablecaption{
Morphological parameters for artificial galaxies.\label{tb:tb1}
}
\tablehead{\colhead{$m$} & \colhead{$r_\mathrm{e}$} & \colhead{$n$} & \colhead{$b/a$} & \colhead{$PA$} \\ 
\colhead{(mag)} & \colhead{(pixel)} & \colhead{} & \colhead{} & \colhead{} } 
\startdata
19--28 & 2--30 & 0.5--10 & 0.1--1.0 & 0--180 \\
\enddata

\end{deluxetable}

%%%%%%%%%%%%%%%%%%%%%%%%%%%%%%%%%%%%%%
\section{Analyses for MODS Galaxies in the GOODS-N Region}\label{sec:sec4}

\subsection{Comparisons with Different PSFs} \label{sec:sec4.1}
\begin{figure*}
\figurenum{3a}
\plotone{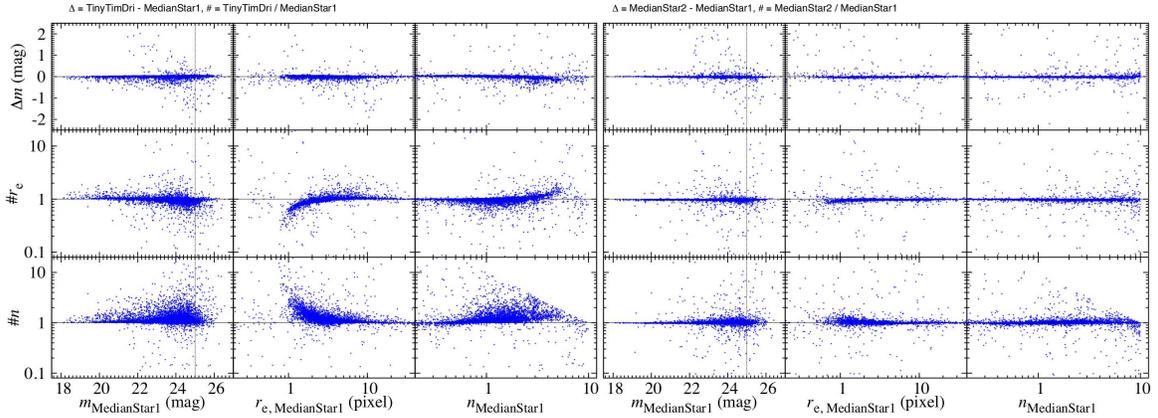}
\caption{
\textcolor{black}
{
{\it Left} : Comparison of the results for $m$, $r_\mathrm{e}$ and $n$ (lower) of {\ttfamily GALFIT} using Drizzled Tiny Tim PSF and median stacked star (MedianStar1) for PSF images. 
$m_\mathrm{MedianStar1}$, $r_\mathrm{e,MedianStar1}$ and $n_\mathrm{MedianStar1}$ are those derived with MedianStar1 (FWHM=0\arcsec\!.18), while $m_\mathrm{TinyTim}$, $r_\mathrm{e,TinyTim}$ and $n_\mathrm{TinyTim}$ are those with the drizzled Tiny Tim PSF.
The ordinate is the difference of the two derived parameters defined as $\Delta m=m_\mathrm{TinyTim}-m_\mathrm{MedianStar1}$, $\#r_\mathrm{e}=r_\mathrm{e,TinyTim}/r_\mathrm{e,MedianStar1}$ and $\#n=n_\mathrm{TinyTim}/n_\mathrm{MedianStar1}$.
{\it Right} : The same comparison as the left panel, but with two median stacked stars, MedianStar1 and MedianStar2 of FWHM=0\arcsec\!.21.
}
}
\label{fig:fig3a}
\end{figure*}

\begin{figure*}[htbp]
\figurenum{3b}
\begin{center}
\plotone{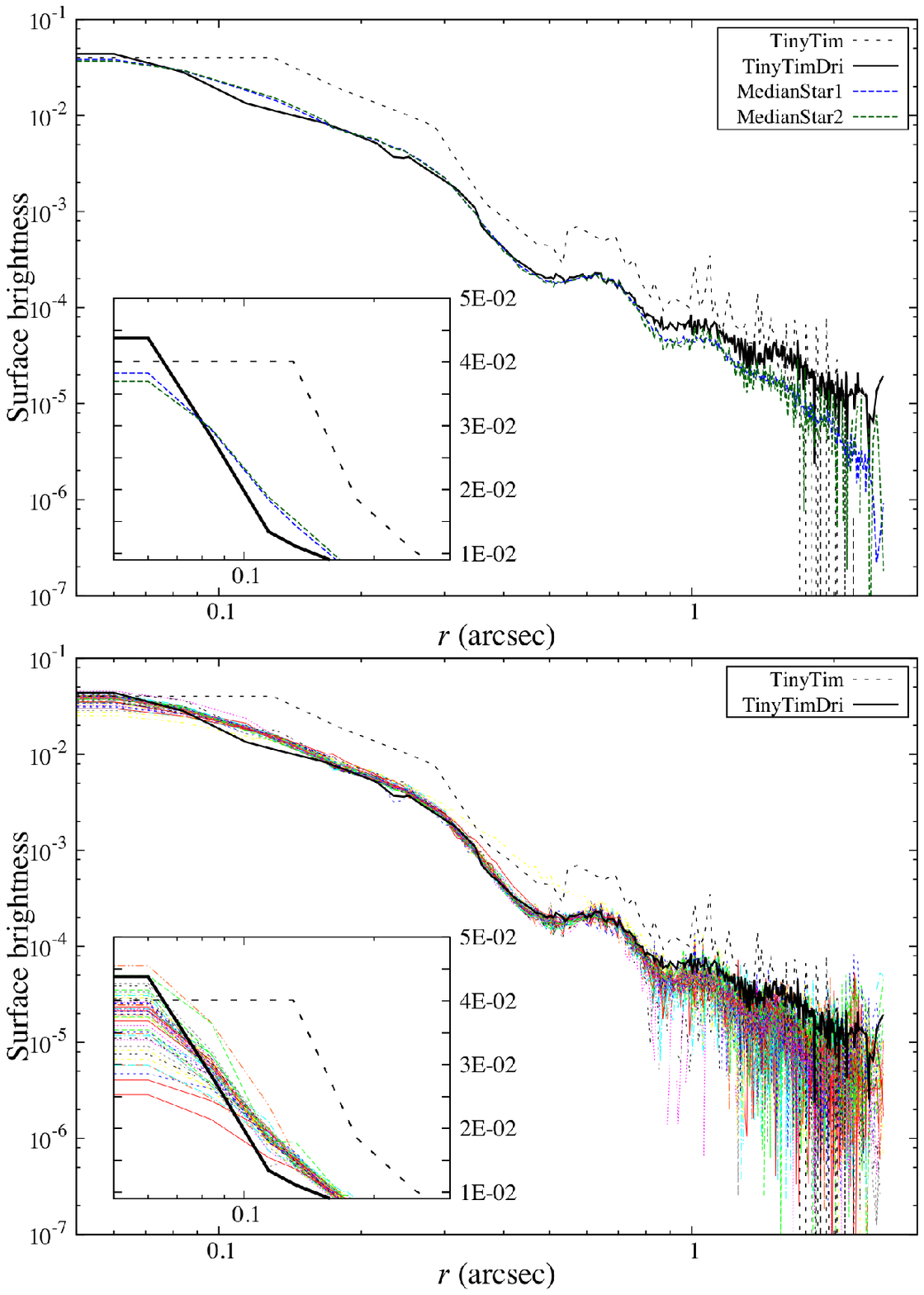}
\caption{
Comparison of the radial profiles of PSFs.
The radial profiles of four PSFs are shown in the upper panel; original Tiny Tim (black dotted line), drizzled Tiny Tim (black solid line), median stacked stars (blue and green dash lines).
The profiles are scaled so as to have same total magnitude.
It is noted that the original Tiny Tim PSF is re-sampled into the same pixel scale as the other PSF.
Inset shows the inner part of the profiles with linear ordinate scale.
The radial profiles of individual stars are also shown in the lower panel.
}
\label{fig:fig3b}
\end{center}
\end{figure*}

A PSF profile is one of the most important parameters to be carefully treated for the morphological fitting by {\ttfamily GALFIT}. 
Since a small change of the PSF profile sometimes leads to incorrect estimates of $r_\mathrm{e}$ and $n$, an appropriate PSF for each image should be used. 
In that context, we estimate how different PSFs affect the best fit morphological parameters for galaxies.
In the previous studies of morphology with {\ttfamily GALFIT}, there have been two stars used; median stacked stars and Tiny Tim PSF.
In order to see the effect, we prepare two PSF images.
Firstly, we stack unsaturated stars in the GOODS-N image to make a PSF image (FWHM $\sim0\arcsec\!.18$).
The sky background of the image is subtracted.
Since the PSFs of stars varies in profiles on the WFC3 detector position, we choose the stars near the detector center. 
(The effect of the PSF change is discussed below.)
Another PSF is Tiny Tim PSF (Krist~\citeyear{krist95}), which simulates PSFs for the $HST$ images. 
Since Tiny Tim PSF is not designed for the drizzled image, we drizzle it in the same manner for the science images.
The self-drizzled Tiny Tim PSF was also used in van~der~Wel et al.~(\citeyear{vanderwel12}), where they replaced the central pixels of the median stacked star by Tiny Tim model PSF to make a hybrid PSF.
The {\ttfamily GALFIT} results with the median stacked PSF and the drizzled Tiny Tim PSF are compared in the left panel of Fig.~\ref{fig:fig3a}.
Although the derived $m$ values show good consistency, $r_\mathrm{e}$ and $n$ show non-negligible difference, especially at smaller $r_\mathrm{e}$ and larger $n$, which also suggests a strong correlation between $r_\mathrm{e}$ and $n$.
Bruce et al.~(\citeyear{bruce12}) found that the Tiny Tim PSF gave galaxy sizes systematically 5--10\% larger than those determined with a median stacked star.
To investigate the origin of the inconsistency, we compare the profiles of those PSFs in Fig.~\ref{fig:fig3b}.
The original Tiny Tim PSF differs from the other PSFs, even though the original Tiny Tim PSF is re-sampled into the pixel scale of the drizzled PSFs.
The Drizzled Tiny Tim PSF and median stacked stars look similar at a glance.
In fact, the FWHM of the drizzled Tiny Tim PSF is $\sim0\arcsec\!.18$, which is in good agreement with those of stacked stars.
However, the inner profile shows non-negligible difference, which may causes the discrepancy in $r_\mathrm{e}$ and $n$ derived with those PSFs.
Therefore, since both of the original and drizzled Tiny Tim PSF do not well simulate the observed PSF, we adopt the median stacked star in the following analysis.
\textcolor{black}{
It is noted that the sub pixel offsets of real PSF stars can broaden the central pixels of the median PSF.
On that point, the median stacked PSF is not necessarily the true PSF either.
} 

FWHMs of unsaturated stars in our $H_\mathrm{160}$ image are found to be in the range of $\sim 0\arcsec\!.18$--$0\arcsec\!.21$, which are sightly larger than those obtained by the ground test of the instruments (Bond et al.~\citeyear{bond07}) and previous studies in other regions (e.g., van~der~Wel et al.~\citeyear{vanderwel12}).
\textcolor{black}{
This could affect the morphological measurements.
}
To investigate whether the variance of PSFs over the WFC3 detector affects the final results, we stack stars on different positions of the detector to make two PSF images with FWHMs $\sim0\arcsec\!.18$ (MedianStar1) and $\sim0\arcsec\!.21$ (MedianStar2).
The results with the PSFs are shown in the right panel of Fig.~\ref{fig:fig3a}, where we see no significant difference, which is consistent with Akiyama et al.~(\citeyear{akiyama08}), who studied S\'ersic profiles derived with PSFs of FWHMs ($\sim0\arcsec\!.13$-$0\arcsec\!.21$) for AO imaged high-$z$ galaxies and found that there was little difference in $r_\mathrm{e}$, though they found non-negligible difference in $n$ ($\Delta n \sim1$).
As such, in what follow we adopt the median stacked stars with FWHM $\sim0\arcsec\!.19$ for the following analysis.
It is noted that the variance of FWHMs of stars are independent of their position in the mosaic images.
This is reasonable for the CANDELS observations because stacked images are shifted and rotated each other, averaging the instrumental aberrations.

%%%%%%%%
% Graphics %
%%%%%%%%

\subsection{Color Effect on Galaxy Sizes} \label{sec:sec4.2}
\begin{figure}
\figurenum{4}
\epsscale{1}
\plotone{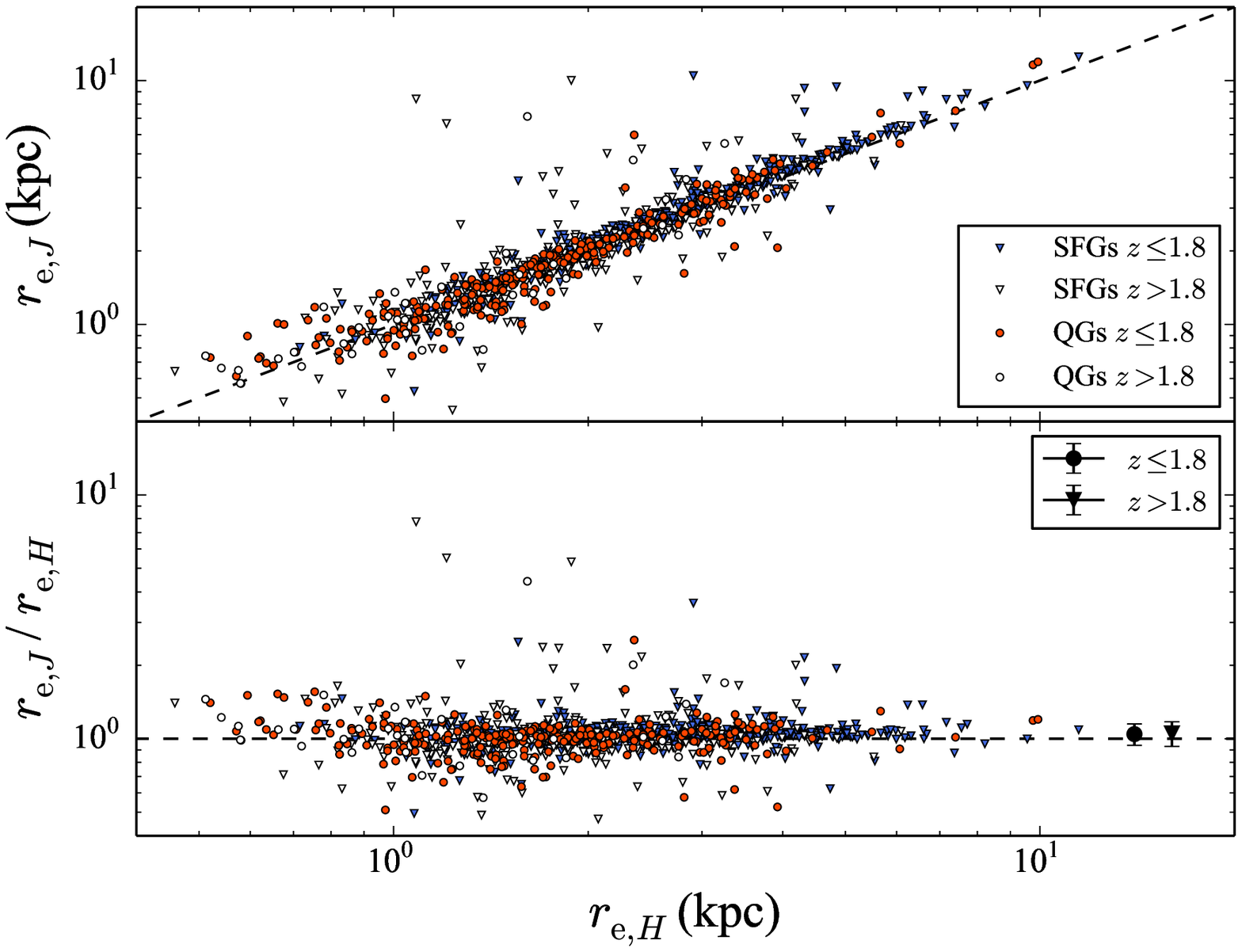}
\caption{
Comparison of $r_\mathrm{e}$ for $J_{125}$ ($r_{\mathrm{e},J}$) and $H_{160}$ ($r_{\mathrm{e},H}$) images. 
Red circles and blue triangles represent QGs and SFGs, respectively.
Filled and open symbols represent galaxies at $z\leq$1.8 and $z>$1.8, respectively.
The median offsets with the MADs for the galaxies at $z\leq1.8$ and $z>1.8$ are shown at the right edge in the bottom panel as filled circle and triangle, respectively.
}
\label{fig:fig.4}
\end{figure}

To examine the color effect on $r_\mathrm{e}$, we compare the sizes of the galaxies obtained on two different images, $J_\mathrm{125}$ and $H_\mathrm{160}$ of WFC3. 
$J_\mathrm{125}$ and $H_\mathrm{160}$ filters correspond to the rest-frame $V$-band image at $z\sim$ 1.0--1.8 and 1.8--3.0, respectively, while the $H_\mathrm{160}$ band corresponds to the rest-frame NIR wavelengths ($\sim$ 0.8--1.2~$\micron$) at $z\sim$ 0.5--1.0.
We examine if there are any significant differences in the sizes obtained at the rest-frame optical and NIR wavelengths. 
In the same manner for AGs in Section~\ref{sec:sec3}, we analyze 1,646 MODS galaxies with $M_*\geq10^{10}M_\mathrm{\sun}$ at $0.5\leq z\leq3.0$ in $J_\mathrm{125}$ and $H_\mathrm{160}$ bands, independently. 
We provide the median stacked stars as a PSF image for each filter, following the previous section. 
The comparison of the results for 1,071 common galaxies successfully fitted with {\ttfamily GALFIT} in both images is shown in Fig.~\ref{fig:fig.4}.
Although there is offset for small galaxies, we see no difference between the radii of the galaxies at $z\leq1.8$ and $z>1.8$.
The focus on one band could avoid extra concerns, such as PSFs and image quality. 
Therefore, in what follows we adopt the $H_\mathrm{160}$ image for the analysis of the galaxies at $z\sim$ 0.5--3.0 in the GOODS-N region.

%%%%%%%%
% Graphics %
%%%%%%%%

\subsection{Stellar Mass Correction for the Galaxies in MODS Catalog} \label{sec:sec4.3}
\begin{figure}
\figurenum{5}
\plotone{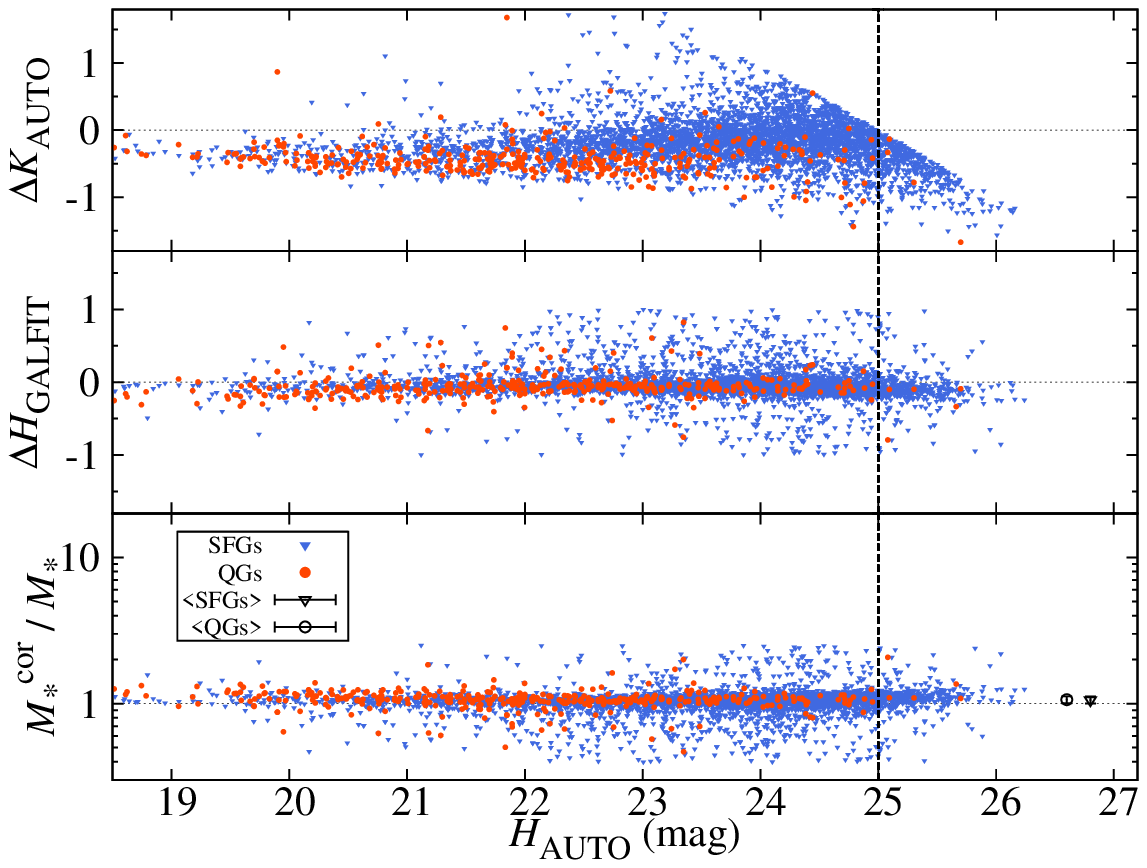}
\caption{
{\it Top} : Comparison of $H_\mathrm{AUTO}$ derived in the present study and $K_\mathrm{AUTO}$ from MODS catalog.
$\Delta K_\mathrm{AUTO}$ is defined as $K_\mathrm{AUTO}-H_\mathrm{AUTO}$.
Filled red circles and filled blue triangles are QGs and SFGs, respectively.
{\it Middle} : Difference between $H_\mathrm{GALFIT}$ and $H_\mathrm{AUTO}$ ($\Delta H_\mathrm{GALFIT}=H_\mathrm{GALFIT}-H_\mathrm{AUTO}$).
{\it Bottom} : Comparison of $M_*^\mathrm{cor}$ and $M_*$ with $H_\mathrm{AUTO}$.
The median offsets for QGs (open circle) and SFGs (open triangle) are shown at the right edge in the bottom panel with their MADs.
}
\label{fig:fig.5}
\end{figure}

Before we obtain the size-stellar mass relations for the present sample, we should correct the stellar mass of each galaxy because the flux measured through non-parametric method would be different from that based on parametric {\ttfamily GALFIT}. 
Since the stellar masses of the MODS galaxies were derived by using non-parametric magnitude of $K_\mathrm{s}$-band images measured by {\ttfamily SExtractor} ($K_\mathrm{AUTO}$), we correct them into those by the model magnitude of {\ttfamily GALFIT}.
The correction is made with a following equation,
\begin{equation}
M_*^\mathrm{cor} = 10^{-0.4(H_\mathrm{GALFIT}-H_\mathrm{AUTO})}M_* ,
\end{equation}
where $H_\mathrm{GALFIT}$ is the model magnitude obtained with {\ttfamily GALFIT} on $H_{160}$ image. 
As MODS catalog does not give $H$-band non-parametric magnitude due to the shallower observation, we use those derived with $H_{160}$ image. 
The small offset due to galaxy colors is irrelevant to magnitude as seen in the upper panel of Fig.~\ref{fig:fig.5}.
In addition, we compare $H_\mathrm{AUTO}$ and $H_\mathrm{GALFIT}$ in the middle panel, where we find no systematic difference between the two magnitudes (non-parametric and parametric).
The comparison of $M_*^\mathrm{cor}$ and $M_*$ are shown in the bottom panel of the figure with median values for QGs and SFGs, where the difference is found to be negligible.
Although the sample selection is made based on $M_*$, the size-stellar mass relations in the following section are obtained using $M_*^\mathrm{cor}$.

%%%%%%%%
% Graphics %
%%%%%%%%

%%%%%%%%%%%%%%%%%%
\section{Results}\label{sec:sec5}
After extracting MODS galaxies from the mosaic images into pixel postage stamps with sides of $r_\mathrm{fit}$, 1,669 massive ($\geq10^{10}$~M$_\mathrm{\sun}$) galaxies at $0.5\leq z\leq3.0$ are analyzed.
We obtain profile parameters for the galaxies in the same manner as done for AGs in Section~3.
From 1,646 galaxies with $M_* > 10^{10} $~M$_{\sun}$ at $0.5\leq z\leq3.0$ listed in MODS catalog, 97 galaxies with neighboring galaxies or stars are discarded to avoid the systematic bias.
We apply {\ttfamily GALFIT} to the remaining galaxies to obtain S\'ersic parameters. 
The results for 39 galaxies with unrealistic morphological profiles, $n < 0.1$, $n > 10$, $r_\mathrm{e} > 60$ pixel ($\sim30$ kpc at $z\sim1.5$), $b/a<0.1$ (see Section~5.1), or inconsistent magnitudes between {\ttfamily GALFIT} and {\ttfamily SExtractor}, $|\Delta m_H| = |H_\mathrm{GALFIT}-H_\mathrm{AUTO}| > 1.0$ are excluded from the resuls.
\textcolor{black}{
In addition, we discard 128 X-ray sources detected by the Chandra Deep Field North (CDF-N) survey (Alexander et al.~\citeyear{alexander03}) to avoid the possible bias due to AGN.}
We finally obtain reliable S\'ersic parameters for 1,382 galaxies.\footnotemark[2]
\footnotetext[2]{The catalog for the morphological parameters with MODS ID can be downloaded from\\
\href{http://www.astr.tohoku.ac.jp/~mtakahiro/}{http://www.astr.tohoku.ac.jp/$\sim$mtakahiro/}.}
It is noted that in what follow we do not include the samples in incomplete redshift and stellar mass bins (see Section~\ref{sec:sec2} and Fig.~\ref{fig:fig.1}) in the statistical discussion, though we show them in the figures.

\subsection{Size-Stellar Mass Relations for Two Galaxy Populations}\label{sec:sec5.1}
\begin{figure}
\figurenum{6}
\plotone{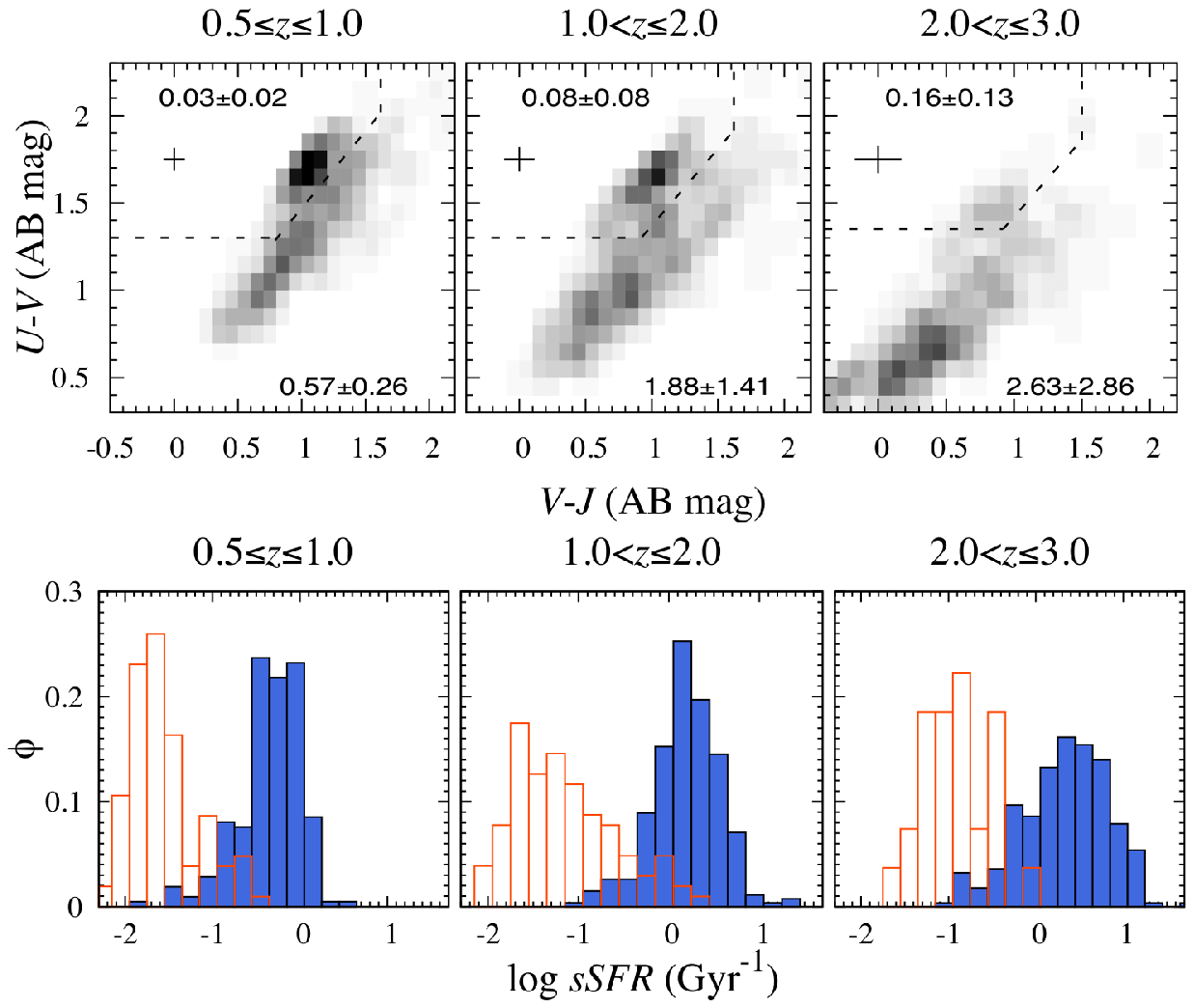}
\caption{
\textcolor{black}{
{\it Top} : Color selections for QGs and SFGs for the same redshift bins in Williams et al.~(\citeyear{williams09}).
Cross symbol represents 1~$\sigma$ error in $U-V$ and $V-J$ for each redshift bin.
Median $sSFRs$ with MADs for QGs and SFGs are written at left top and right bottom, respectively, in each panel.
{\it Bottom} : Histograms of QGs (red open) and SFGs (blue filled) as a function of $sSFR$.
The ordinates for QGs and SFGs are normalized by the total number of each population in the redshift bins, respectively.
}
}
\label{fig:fig.6}
\end{figure}

Before we derive the size-stellar mass relations, we separate the present sample into QGs and SFGs in upper panel of Fig.~\ref{fig:fig.6}, following the color selection criteria described by Williams et al.~(\citeyear{williams09}), which used rest-frame $U-V$ and $V-J$ colors as follows;
\begin{eqnarray}
(U-V) > 0.88(V-J) + c, 
\end{eqnarray}
where $U-V$ and $V-J$ in the rest frame were obtained with the SED-model fit to galaxies. 
The offset, $c$, is 0.59 and 0.49 for $0.5 < z < 1.0$ and $1.0 < z < 2.0$, respectively. 
Additional criteria of $U-V>1.3$ and $V-J<1.6$ are imposed on QGs at all redshifts to exclude obscured and dusty SFGs, respectively.
For $z > 2.0$ galaxies, Williams et al.~(\citeyear{williams09}) concluded that there were no visible two sequence and applied the offset for $1.0 < z < 2.0$, which would be less reliable.
As such, we modified the criteria for QGs and SFGs at $z\geq2$.
To see the bimodality, we use specific star-formation rate ($sSFR$), which is derived from UV and IR-luminosity (Kennicutt~\citeyear{kennicutt98}; see also Kajisawa et al.~\citeyear{kajisawa10}).
In the lower panel of Fig.~\ref{fig:fig.6} we show the histograms of the two population as a function of $sSFR$, where we can see clear bimodality even at $z\geq2.0$.
We adopt the criteria for QGs at $z\geq2.0$ so that the overlap of two population on $sSFR$ become minimum;
\begin{center}
\begin{eqnarray}
(U-V) > 0.88(V-J) + 0.54 \nonumber \\
\cap (U-V) > 1.35 \\
\cap (V-J)<1.50\nonumber.
\end{eqnarray}
\end{center}
It is noted that the average values of $sSFR$ raise as redshift increases (e.g., Daddi et al.~\citeyear{daddi07}; Peng et al.~\citeyear{peng10}), and adopting fixed $sSFR$ for the selection of QGs is not appropriate for our purpose.
\textcolor{black}{
It should be also noted that adopting different criteria for $UVJ$ color selection (e.g., Whitaker et al.~\citeyear{whitaker11}) would not change our final results.}
%%%%%%%%
% Graphics %
%%%%%%%%

Moreover, 24 QGs detected at 24$\micron$ by $Spitzer$/MIPS are included in SFGs.
Our results are hardly changed even if we discard the dust-obscured SFGs.
Based on the color selection, we obtain 299 QGs and 1,083 SFGs.
The redshift bins are defined as in Ic12 to facilitate the comparison.
We also exclude \textcolor{black}{24 galaxies} with $b/a<0.1$, because real galaxies have $b/a>0.1$ (e.g., Binney \& de Vauclouleurs~\citeyear{binney81}).
Since the single component S\'ersic fit by {\ttfamily GALFIT} in the present study does not take account of the inclination, the surface brightness profile of inclined spheroidal or disk galaxies may not be presented by the profile.
To derive the exact light profile for inclined galaxies, we need to calculate the integrated light profile combined with the opacity for each wavelength (e.g., Graham \& Worley \citeyear{graham08}), which is beyond the present scope.

\begin{figure*}
\figurenum{7}
\plottwo{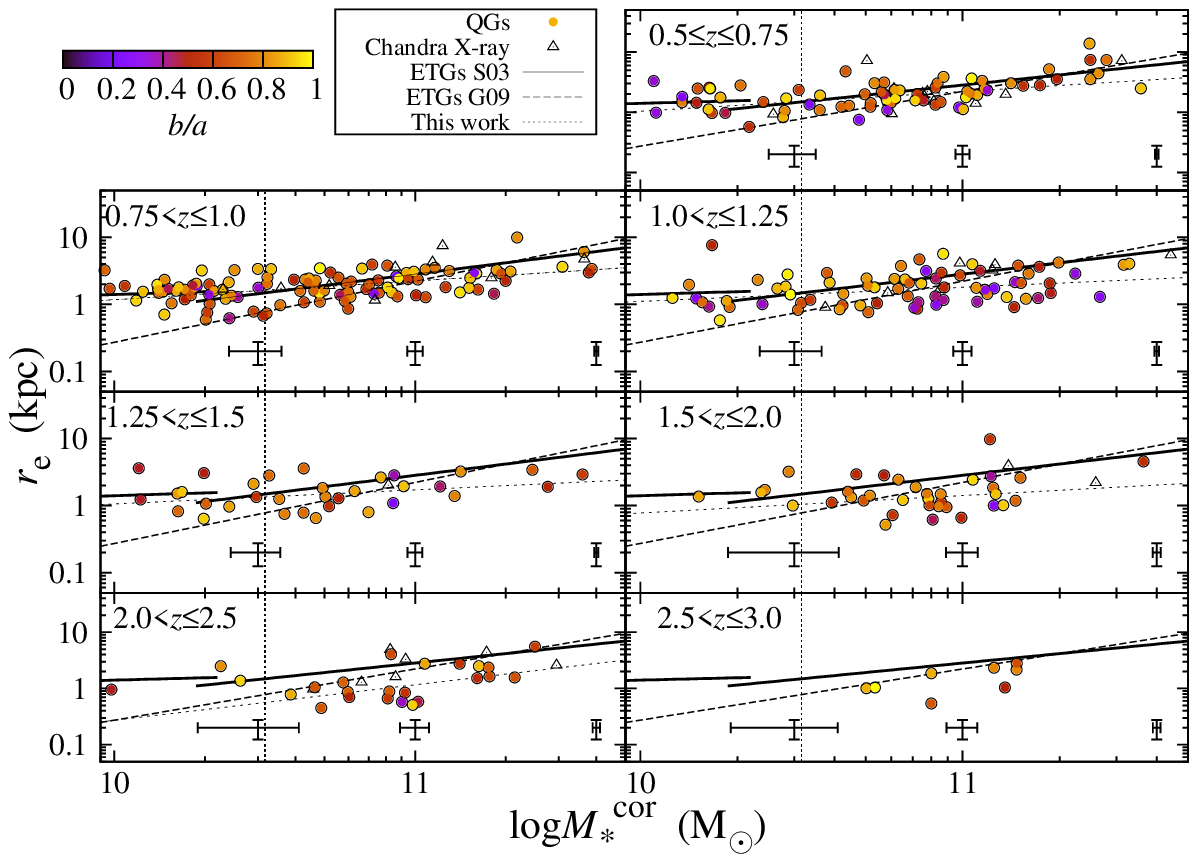}{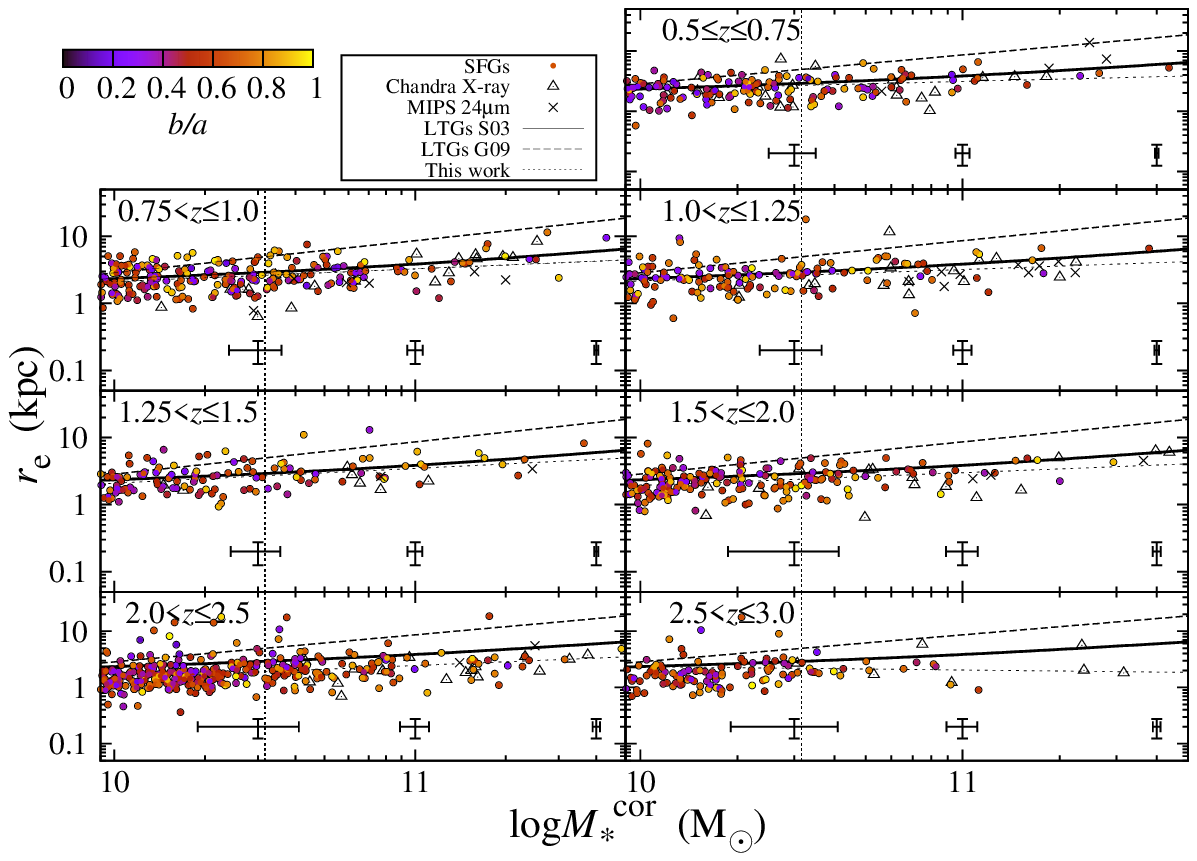}
\caption{
{\it Left} : Size-stellar mass relations for effective radius, $r_\mathrm{e}$, of QGs in the GOODS-N (filled circles). 
The color banner shows the $b/a$ value.
The local size-stellar mass relations for early-type galaxies of S03 (solid line) and G09 (dashed) are depicted. 
The relation for the present samples in each redshift bin is also shown (dotted line).
X-ray detected sources by Chandra (open triangles) are not included in the regressions.
Error bars near the bottom of each redshift panel show the MADs of the stellar mass and $r_\mathrm{e}$ for each redshift bin.
It is noted that the MAD for size represents the maximum error derived by {\ttfamily GALFIT} analysis with $n>2.5$ at $24<H_\mathrm{AUTO}\leq25$ in Fig.~\ref{fig:fig.2b}.
The vertical dotted line for each redshift bin is the boundary of $M_*^\mathrm{cor}=10^{10.5} $~M$_{\sun}$.
{\it Right} : Same as the left panel but for SFGs.
QGs detected by Spitzer/MIPS 24$\micron$ are represented by crosses.
The local size-stellar mass relations for late-type galaxies of S03 (solid) and G09 (dashed) are depicted.
}
\label{fig:fig7}
\end{figure*}

The size-stellar mass relations are plotted in Figs.~\ref{fig:fig7} for QGs and SFGs, respectively. 
S03 obtained the relations for ETGs and late-type galaxies (LTGs) in the local universe, which is often compared with those at high-$z$. 
In addition, we plot the relations for the central galaxies (CENs) of galaxy groups and clusters in the SDSS by G09, though most of the galaxies in the present study are in field environment. 
G09 reanalyzed the SDSS galaxies with {\ttfamily GALFIT} and derived the relation of the stellar mass with $r_\mathrm{e}$, whereas S03 was based on a non-parametric radius.
As noted in Section~\ref{sec:sec6}, the non-parametric radius is systematically different from parametric $r_\mathrm{e}$.
The stellar masses for SDSS data are converted to those which would be obtained with BC03 models (Bruzual \& Charlot~\citeyear{bruzual03}) and Salpeter IMF using the relations in Cimatti et al.~(\citeyear{cimatti08}).

%%%%%%%%
% Graphics %
%%%%%%%%

\begin{figure}
\figurenum{8}
\plotone{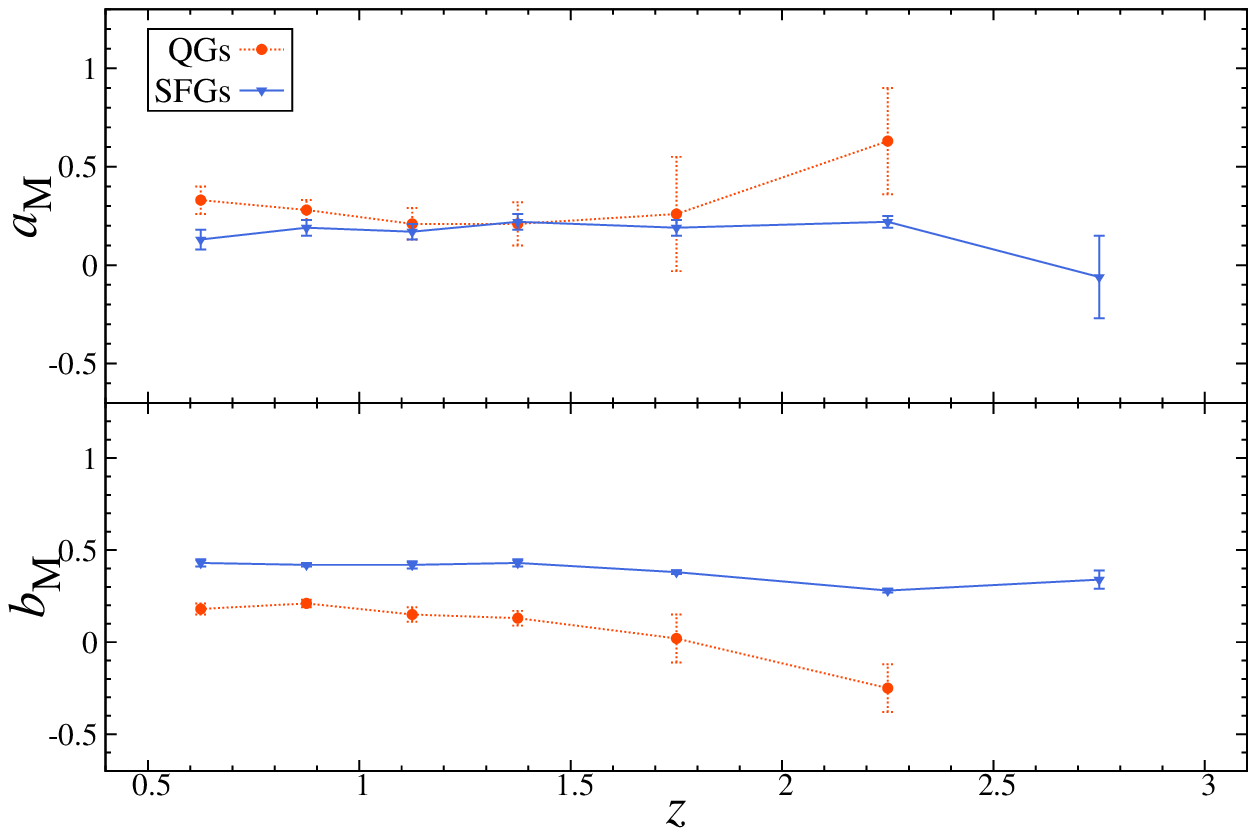}
\caption{
Best fit slopes ($a_\mathrm{M}$, {\it top}) and offsets ($b_\mathrm{M}$, {\it bottom}) for the size-stellar mass relations as a function of redshift for QGs (red solid lines) and SFGs (blue dotted lines).
Bars represent mean errors.
}
\label{fig:fig.8}
\end{figure}

It is clear in Figs.~\ref{fig:fig7} that the sizes of QGs and massive SFGs at higher redshift are smaller than those of the galaxies in the local universe at a given mass.
There are also some massive compact galaxies whose sizes are a factor of $\sim5$ smaller than those of typical local ETGs. 
The sizes of the compact galaxies are comparable to those found by the previous studies. 
To see the evolution of the size-stellar mass relation, we then obtain the least square fit with a following linear regression,
\begin{equation}
logr_\mathrm{e}=a_\mathrm{M}\ log(M_*^\mathrm{cor}/M_c)+b_\mathrm{M},
\end{equation}
where $M_\mathrm{c}$ is the characteristic mass for the massive galaxies, here set to $10^{10.5}$~M$_{\sun}$. 
The best fit slope, $a_\mathrm{M}$, and offset, $b_\mathrm{M}$, are shown in Fig.~\ref{fig:fig.8} for QGs and SFGs.
$a_\mathrm{M}$ for the SFGs remains unchanged over the redshifts.
For QGs at higher redshift ($z\geq2.0$),  $a_\mathrm{M}$ is slightly higher value with the large error because the number of the sample is very small and incomplete.
The slope is consistent \textcolor{black}{within the error} with those of previous studies (e.g., Williams et al.~\citeyear{williams10}; Cimatti et al.~\citeyear{cimatti12}).

%%%%%%%%
% Graphics %
%%%%%%%%

\subsection{Evolution of Galaxy Size with Redshift} \label{sec:sec5.2}

\begin{figure*}
\figurenum{9}
\epsscale{1.0}
\plotone{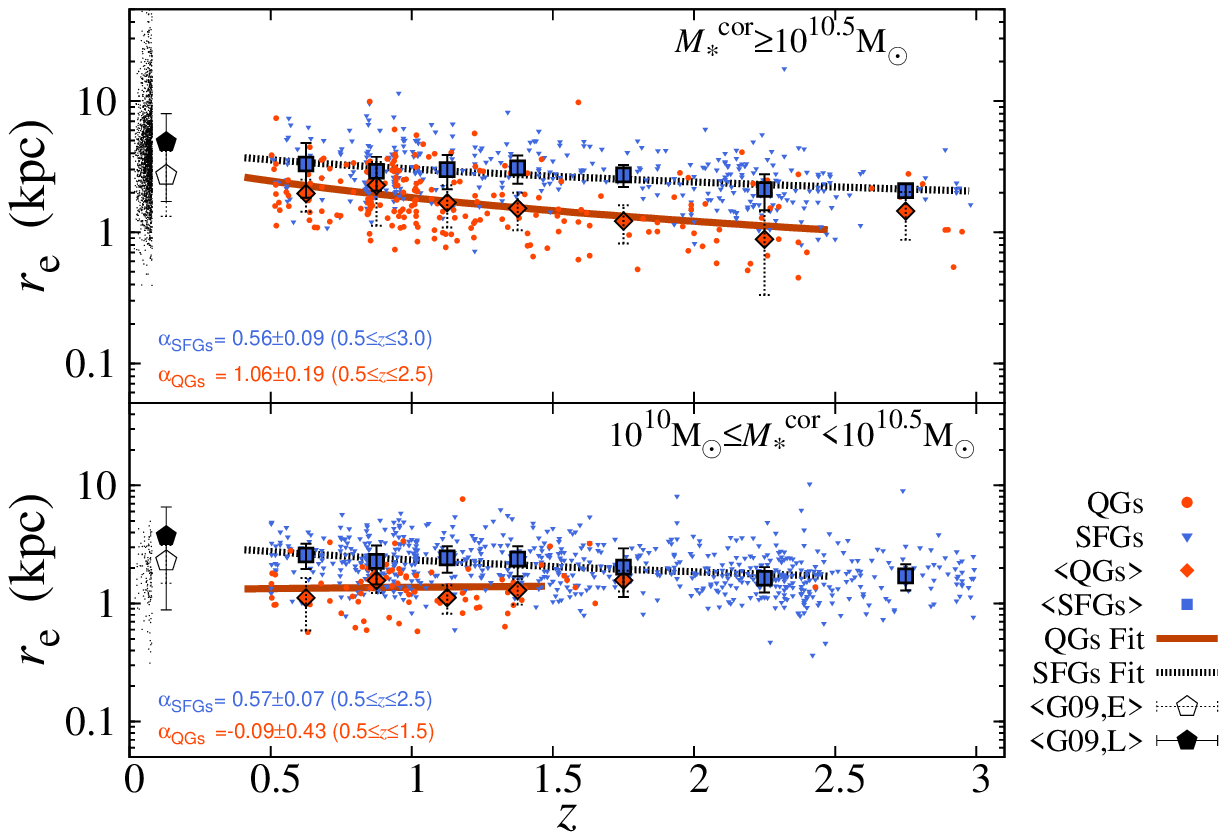}
\caption{
Evolution of $r_\mathrm{e}$ as a function of redshift for massive ($M_*^\mathrm{cor}\geq10^{10.5}$~M$_{\sun}$, $top$) and less massive ($M_*^\mathrm{cor}<10^{10.5}$~M$_{\sun}$, $bottom$) QGs (red circles) and SFGs (blue triangles).
The median sizes with MADs for QGs (red filled diamonds with dotted bars) and SFGs (blue filled squares with solid bars) are also shown.
The local galaxies of CEN samples are also plotted (black dots) with the median sizes for early-type ($n>2.5$, black filled pentagons) and late-type ($n\leq2.5$, black open pentagons) galaxies. 
The thick solid lines and break lines represent the regression for QGs and SFGs, respectively.
It is noted that the samples in the incomplete redshift bins are not included in the regression.
The median sizes and best fit coefficients are summarized in Table~3 and 4, respectively.
}
\label{fig:fig.9}
\end{figure*}

We show the size evolution of the present sample as a function of redshift in Fig.~\ref{fig:fig.9}. 
In Table~3, we summarize the median size, $\langle{r_\mathrm{e}}\rangle$, and the median absolute deviation (MAD) for each redshift bin. 
To see the dependence of the evolution on the stellar mass, we divide the sample into the massive and less massive groups as above.
We plot the local sample of early-type ($n>2.5$) and late-type ($n\leq2.5$) CENs and their median sizes to compare with our results; the stellar masses for CENs sample are converted into those of Salpeter IMF as done in the previous section.
As shown in the figure, the median size for the massive QGs is smaller than that of the local galaxies, which evolves to the size of local galaxies over the redshifts.
(The larger median size of the sample at the highest redshift bin of $2.5<z\leq3.0$ is less reliable due to the small sample number.)

We derived the regressions for the sizes of galaxies with $r_\mathrm{e} \propto (1 + z)^{-\alpha_{r_\mathrm{e}}}$, and obtained $\alpha_{r_\mathrm{e}} =1.06\pm0.19$ for massive QGs at $0.5\leq z\leq2.5$ (increase by a factor of $\sim2.5$ in size over the redshift) and $\alpha_{r_\mathrm{e}} =0.56\pm0.09$ ($\sim1.7$) for massive SFGs at $0.5\leq z\leq3.0$.
For less massive QGs, the sizes seem to be unchanged over the redshifts, though they are not statistically robust due to the small sample number and difference in the range of redshift.
It is noted that the local SDSS galaxies are not included in the regression.
The best fit values with their errors are summarized in Table~4.

Ic12 showed the weak size evolution in half-light radius ($r_{50}$) for the present sample based on the ground-based images in $K_\mathrm{s}$-band.
The result of the weak size evolution of Ic12 (a factor $\lesssim1.3$ from $z\sim2.5$ to $z\sim0.3$) contradicts with the significant size evolution of the present study and many other previous ones which presented the size evolutions by a factor of $\sim$ 2--5 for the similar redshift ranges.
We investigate the inconsistency of the present results with Ic12 in Appendix~\ref{sec:app}, mainly focusing on the two different method of size measurement of $r_\mathrm{e}$ and $r_{50}$.

Although Newman et al.~(\citeyear{newman10}) obtained a relatively weaker evolution ($\alpha_{r_\mathrm{e}} \sim0.75\pm0.10$) for massive ($M_\mathrm{den}>10^{11}$~M$_{\sun}$) field spheroid over $0 < z < 1.6$, most of other previous study based on {\ttfamily GALFIT} have presented the significant evolution for QGs (Franx et al.~\citeyear{franx08} $\alpha_{r_\mathrm{e}}\sim1.09\pm0.07$ for $M_*\geq 4\times10^{10}$~M$_{\sun}$ QGs at $0.5\leq z\leq 3.5$; Cimatti et al.~\citeyear{cimatti12} $\alpha_{r_\mathrm{e}} \sim1.24\pm0.15$ for ETGs at $0 < z < 3.0$).
Although Damjanov et al.~(\citeyear{damjanov11}) obtained the highest value of $\alpha_{r_\mathrm{e}} \sim1.62\pm0.34$ for $M_*\geq 10^{10}$~M$_{\sun}$ QGs at $0.2 < z < 2.7$, they obtained the regression for the sizes normalized by stellar masses ($r_\mathrm{e}/{M_*}^{0.51}$), and their $\alpha_\mathrm{e}$ should not be compared with ours.

SFGs in the present study weakly evolve their sizes for both mass bins, which is inconsistent with the previous studies (e.g., Williams et al.~\citeyear{williams10} $\alpha_{r_\mathrm{e}}\sim0.77\pm0.08$ and $\alpha_{r_\mathrm{e}}\sim1.32\pm0.15$ for less massive and massive SFGs, respectively, at $z<2.0$; Mosleh et al.~\citeyear{mosleh11} $\alpha_{r_\mathrm{e}}\sim1.11\pm0.13$ for UV bright galaxies at $0.5\leq z\leq3.5$).
However, there are some differences in analysis between the present study and the previous two studies.
Firstly, Williams et al.~(\citeyear{williams10}) used imaging data obtained by a ground-based telescope.
As discussed in Konishi et al.~(\citeyear{konishi11}), the morphological profiles for smaller galaxies were significantly affected by seeing (FWHM$\sim 0\arcsec\!.5$ for their $K_\mathrm{s}$-band imaging), which made them unable to obtain the reliable S\'ersic profile for the galaxies at $z>1$.
It is also noted that Williams et al.~(\citeyear{williams10}) constrained the output S\'ersic index within $1<n<4$, which could affect the output $r_\mathrm{e}$ because they are strongly correlated (see Fig.~\ref{fig:fig3a}, for example), while we let $n$ down to 0.1 and up to 10, and this may cause the inconsistency in $\alpha_\mathrm{e}$ of SFGs.
It turned out that, however, our results for SFGs hardly changed when we constrained the output S\'ersic index within $1<n<4$.
In Mosleh et al.~(\citeyear{mosleh11}), they used MODS $K_\mathrm{s}$-band image for the size estimation of the $UV$-bright galaxies at $0.5\leq z\leq3.5$, whereas most of our sample consist of normal SFGs.
In addition, their $UV$-bright galaxies consisted of amorphous and irregular galaxies such as submillemeter and Lyman-break galaxies, which have been excluded in our analysis (see Section~\ref{sec:sec3}), and their results could not be directly compared with ours.

\begin{deluxetable*}{lccccccc}
\setlength{\tabcolsep}{0.02in} 
\tabletypesize{\tiny}
\tablewidth{0pt}
\tablenum{3}
\tablecolumns{8}
\tablecaption{
Median sizes and S\'ersic indices with MAD of the galaxies in the GOODS-N region.\label{tb:tb3}
}
\tablehead{
\colhead{} &  \colhead{} & \colhead{}  & \colhead{QGs} & \colhead{} & \colhead{} & \colhead{SFGs}& \colhead{}\\
\cline{3-8}
\colhead{Redshift} &  \colhead{Stellar Mass} & \colhead{N} & \colhead{$<r_{e}>$} & \colhead{$<n>$} & \colhead{N} & \colhead{$<r_{e}>$} & \colhead{$<n>$}\\
\colhead{} &  \colhead{($M_{\sun}$)} & \colhead{} & \colhead{(kpc)} & \colhead{} & \colhead{} & \colhead{{(kpc)}} & \colhead{}}
\startdata
$0.5\leq  z\leq0.75$&$\geq10^{10.5}$	&	36	&	$1.97\pm0.54$	&	$3.78\pm0.85$	& 33	&	$3.31\pm1.48$	&	$1.34\pm0.68$	\\
&$<10^{10.5}$	&	13	&	$1.12\pm0.52$	&	$2.14\pm0.86$	&	52	&	$2.58\pm0.62$	&	$1.19\pm0.26$	\\
$0.75<z\leq1.0$&$\geq10^{10.5}$	&	61	&	$2.27\pm1.16$	&	$3.04\pm1.07$	&	80	&	$2.91\pm0.84$	&	$1.44\pm0.43$	\\
&$<10^{10.5}$	&	34	&	$1.56\pm0.34$	&	$1.95\pm1.15$	&	85	&	$2.27\pm0.82$	&	$1.14\pm0.57$	\\
$1.0<z\leq1.25$&$\geq10^{10.5}$	&	48	&	$1.68\pm0.60$	&	$2.33\pm0.90$	&	48	&	$3.00\pm0.88$	&	$1.48\pm0.45$	\\
&$<10^{10.5}$	&	15	&	$1.12\pm0.30$	&	$1.67\pm0.48$	&	63	&	$2.43\pm0.62$	&	$1.11\pm0.53$	\\
$1.25<z\leq1.5$&$\geq10^{10.5}$	&	20	&	$1.52\pm0.48$	&	$1.88\pm0.80$	&	34	&	$3.09\pm0.75$	&	$1.00\pm0.30$	\\
&$<10^{10.5}$	&	10	&	$1.29\pm0.31$	&	$2.12\pm1.06$	&	49	&	$2.38\pm0.68$	&	$0.87\pm0.44$	\\
$1.5<z\leq2.0$&$\geq10^{10.5}$	&	27	&	$1.21\pm0.39$	&	$1.46\pm1.12$	&	52	&	$2.73\pm0.52$	&	$1.19\pm0.42$	\\
&$<10^{10.5}$	&	5	&	$\--$	&	$\--$	&	72	&	$2.03\pm0.90$	&	$1.07\pm0.48$	\\
$2.0<z\leq2.5$&$\geq10^{10.5}$	&	19	&	$0.88\pm0.55$	&	$0.82\pm0.36$	&	110	&	$2.12\pm0.65$	&	$1.17\pm0.47$	\\
&$<10^{10.5}$	&	1	&	$\--$	&	$\--$	&	155	&	$1.63\pm0.40$	&	$1.07\pm0.59$	\\
$2.5<z\leq3.0$&$\geq10^{10.5}$	&	8	&	$\--$	&	$\--$	&	16	&	$2.07\pm0.25$	&	$1.72\pm1.63$	\\
&$<10^{10.5}$	&	0	&	$\--$	&	$\--$	&	47	&	$\--$	&	$\--$	\\
\enddata

\end{deluxetable*}
%%%
%%%

\begin{deluxetable}{ccccccccc}

\setlength{\tabcolsep}{0.02in} 
\tabletypesize{\tiny}
\tablewidth{0pt}
\tablecolumns{5}
\tablenum{4}
\tablecaption{
Best fit slope for $r_\mathrm{e}$ ($\alpha_{r_\mathrm{e}}$) and $n$ ($\alpha_{n}$) of the galaxies in the GOODS-N region.
\label{tb:tb4}
}
\tablehead{
\colhead{Stellar Mass} &\colhead{QGs}& \colhead{SFGs} & \colhead{QGs}& \colhead{SFGs}\\
\colhead{($M_\odot$)} & \colhead{$\alpha_{r_\mathrm{e}}$} & \colhead{$\alpha_{r_\mathrm{e}}$}&\colhead{$\alpha_n$}  & \colhead{$\alpha_n$}
}
\startdata
$\geq10^{10.5}$&1.06$\pm0.19$&0.56$\pm$0.09 &0.74$\pm0.23$&0.10$\pm$0.13\\
$<10^{10.5}$	&-0.09$\pm0.43$&0.57$\pm0.07$&0.17$\pm$0.21&0.04$\pm$0.09\\
\enddata

\tablecomments{Sample in the incomplete redshift and stellar mass bins are not included in the regressions.}

\end{deluxetable}

%%%%%%%%
% Graphics %
%%%%%%%%

\subsection{Evolution of Surface Density Profiles with Redshift} \label{sec:sec5.3}

\begin{figure*}
\figurenum{10}
\plotone{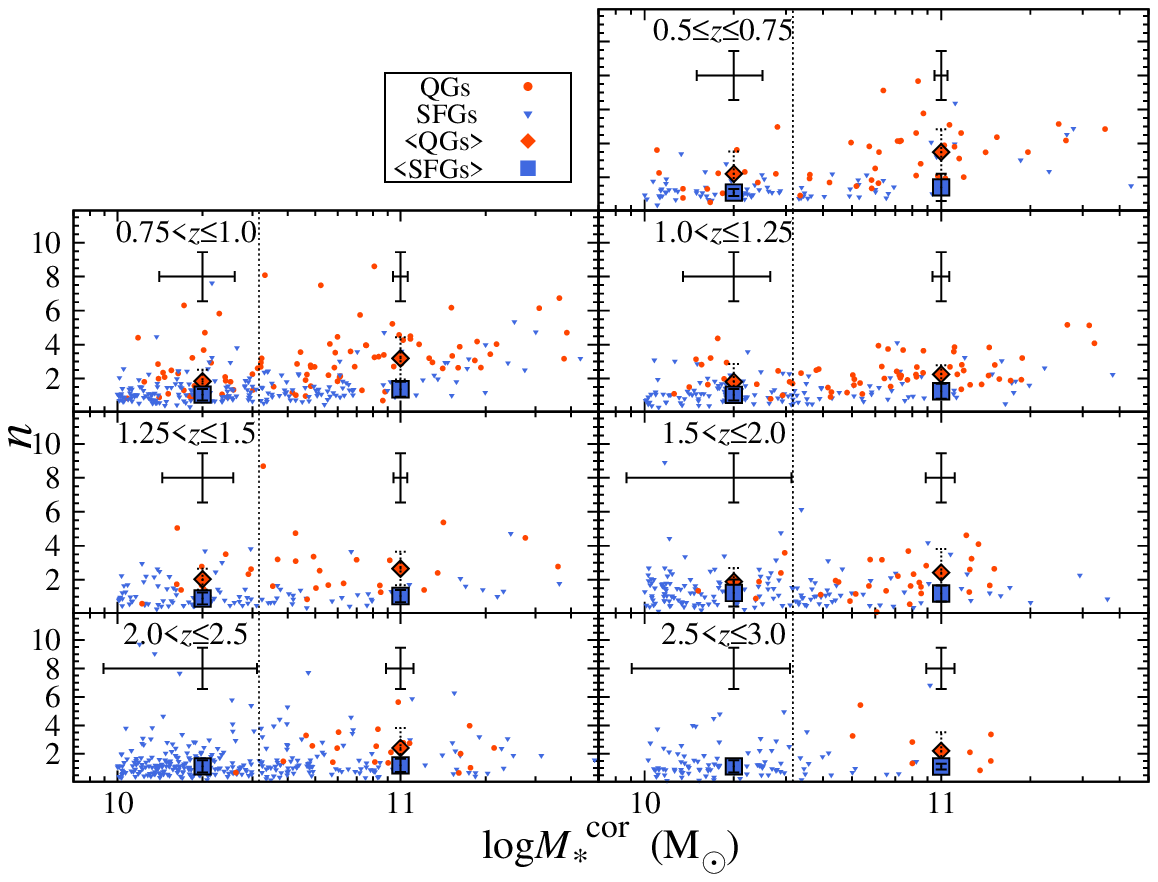}
\caption{
Distribution of S\'ersic index, $n$, with stellar mass for the QGs (red filled circles) and SFGs (blue filled triangles).
The median values with MADs in each redshift bin are shown for QGs (red filled diamonds with dotted lines) and SFGs (blue filled squares with solid lines) with ${M_*}^\mathrm{cor}\geq10.5$~M$_{\sun}$ and ${M_*}^\mathrm{cor}<10.5$~M$_{\sun}$, respectively.
Error bars near the top of each redshift panel represent MADs of the stellar mass (horizontal) and the maximum error for $n$ (vertical) derived by {\ttfamily GALFIT} analysis with $n>2.5$ at $24<H_\mathrm{AUTO}\leq25$ in Fig.~2b.
The vertical dotted line for each redshift bin is the boundary of $M_*^\mathrm{cor}=10^{10.5}$~M$_{\sun}$.
}
\label{fig:fig.10}
\end{figure*}

\begin{figure*}
\figurenum{11}
\plotone{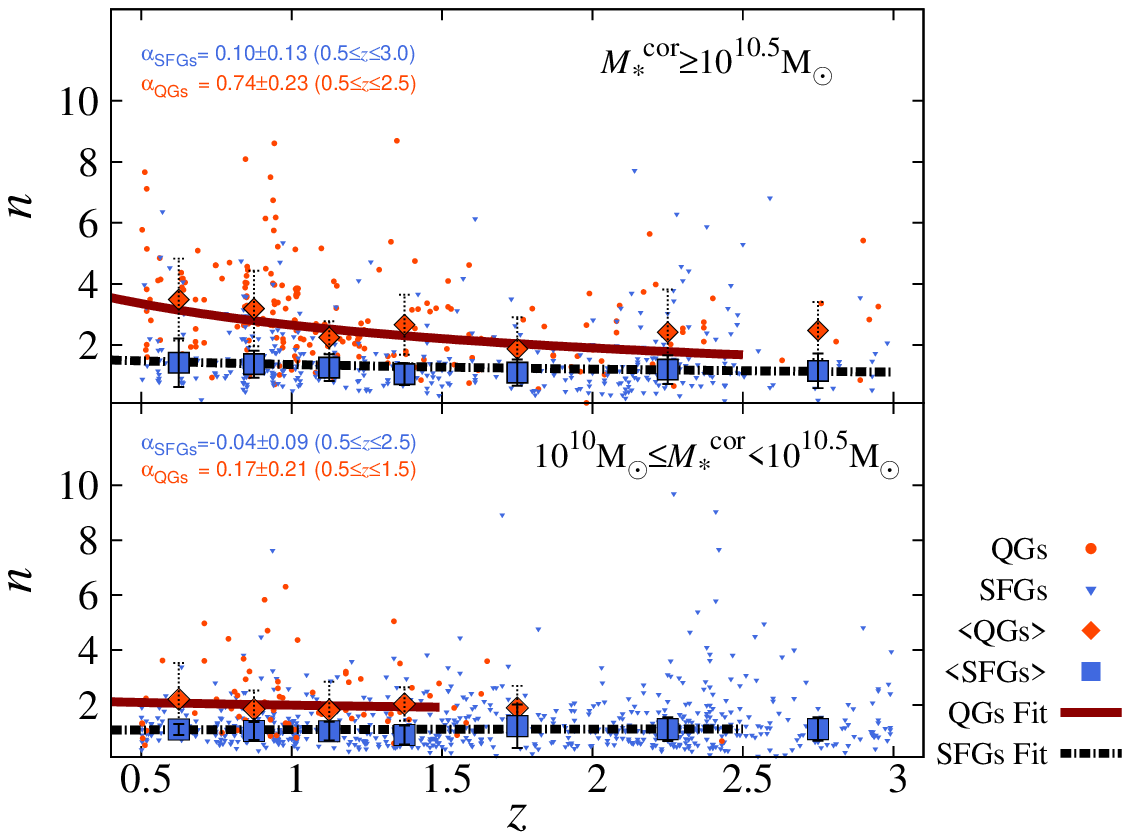}
\caption{
Same as Fig.~\ref{fig:fig.9}, but for $n$.}
\label{fig:fig.11}
\end{figure*}

As we see the size evolution of the galaxies, we here study the evolution of S\'ersic index $n$ in $0.5\leq z\leq 3.0$.
Since $n$ represents the shape of galaxies ($n\sim1$ for disk galaxies, $n\sim4$ for ellipticals), its evolution would give us clues to the formation of local galaxies.
In Fig.~\ref{fig:fig.10}, we show the distribution of $n$ for each redshift bin.
The figure reads the correlations among stellar mass, $n$, and galaxy types (QG and SFG).
On average, SFGs have constant $n$ ($\sim1$) over the redshifts, irrespective of their stellar masses, while QGs have larger value, though $n$ for both groups are widely distributed.
Interestingly enough, we see different evolution of $n$ of the massive and less massive QGs; significant evolution of $n$ of the massive QGs and no evolution of the less massive.
To confirm the trend, we derive the regression for $n$ in Fig.~\ref{fig:fig.11} with $n\propto (1+z)^{-\alpha_n}$ for both groups.
In the figure, we see that median $n$ for SFGs seems unchanged ($\alpha_n\sim0.04$--$0.1$) over $z\sim$ 0.5--2.5, while that of $n$ of massive QGs significantly increases as redshift decreases ($\alpha_n\sim0.74$), except for the highest redshift bin with the significant error.
\textcolor{black}{
The studies for $n$ were also done by van~Dokkum et al.~(\citeyear{vandokkum10}) ($\alpha_n=0.95\pm0.015$) and Patel et al.~(\citeyear{patel13}) ($\alpha_n=0.9\pm0.1$).
However, both of the studies selected the sample irrespective of the populations (QGs and SFGs) of galaxies.
In addition, the sample were selected based on a constant cumulative number density, in which the stellar mass of galaxies increases as redshift decreases.
Therefore, the direct comparison with those studies would be difficult.
}
The different evolution of $n$ depending on the galaxy types and stellar masses should be stressed, especially in the view of the evolution scenario, which is discussed in the following section.
The best fit $\alpha_n$ with their errors are summarized in Table~4.

%%%%%%%%
% Graphics %
%%%%%%%%

%%%%%%%%%%%%%%%%%%%%%%%%%%%%

\section{Discussion} \label{sec:sec6}

We have obtained the size and shape of the massive galaxies, using MODS and $HST$/WFC3 CANDELS data in the GOODS-N region.
Thanks to the high image quality and depth, we were allowed to analyze as faint objects as with $H_\mathrm{AUTO}\leq25$, 
which reaches less massive ($\sim10^{10}$~M$_{\sun}$) SFGs at $z\sim2.5$ (QGs at $z\sim1.5$) and massive ($\geq10^{10.5}$~M$_{\sun}$) SFGs at $z\sim3.0$ (QGs at $z\sim2.5$).
With a careful test of {\ttfamily GALFIT} analysis, applying magnitude criteria to AGs, we obtained unbiased morphological results for the samples.
The tests for different PSFs and color effect also improved the reliability.

First, we discuss the evolution of $r_\mathrm{e}$.
As shown in Fig.~\ref{fig:fig.6} and Fig.~\ref{fig:fig.8}, 
\textcolor{black}{
we found a number of compact QGs ($r_\mathrm{e}\sim1$~kpc or less) at $1.5 < z \leq 2.5$, which is consistent with the previous studies (van~Dokkum et al.~\citeyear{vandokkum09}; Szomoru et al.~\citeyear{szomoru12}).
}
The size for massive QGs from $z\sim2.5$ to $\sim0.5$ is represented as $r_\mathrm{e}\propto (1+z)^{-\alpha_{r_\mathrm{e}}}$ with $\alpha_{r_\mathrm{e}} \sim1.06$ (or a factor of $\sim2.5$ increase from $z\sim2.5$ to $\sim0.5$ at a given stellar mass), which is consistent with the previous results of $\alpha_{r_\mathrm{e}} \sim$ 0.7--1.5 (a factor of $\sim$ 1.8--3.6 size increase).
It is noted that SFGs have weaker size evolution ($\alpha_{r_\mathrm{e}} \sim0.5$), irrespective of their stellar mass bins, which is inconsistent with previous studies (e.g., Mosleh et al.~\citeyear{mosleh11}).
However, the sample definition of SFGs in the studies are different ($UV$-bright galaxies in Mosleh et al.), and their results could not be compared with the present results.
The scenario of the size evolution for QGs is in dispute for a decade both in theoretical (e.g., Fan et al.~\citeyear{fan08}; Hopkins et al.~\citeyear{hopkins09a}; Naab et al.~\citeyear{naab09}; Oser et al.~\citeyear{oser12}) and observational studies (Newman et al.~\citeyear{newman10}; van~Dokkum et al.~\citeyear{vandokkum10}).
According to Eq.~4 in Naab et al.~(\citeyear{naab09}), it takes $\sim4~(\sim2)$ minor merger events with 1:10 (3:10) mass ratio to explain the size evolution of the massive QGs (see also Bezanson et al.~\citeyear{bezanson09}).
In a merger scenario, however, about 10\% of the massive galaxies at $z\sim2$ are expected to have survived without 
\textcolor{black}{
equal-mass merging (Hopkins et al.~\citeyear{hopkins09a}), which leaves the superdense QGs morphologically unchanged to the local universe.
}
In the SDSS and other surveys for the local galaxies, however, there is not enough number of such a superdense relic (S03; Taylor et al.~\citeyear{taylor10}; M\'armol-Queralt\'o et al.~\citeyear{marmol12}).
These facts suggest that the observed size evolution is hard to be explained only by the merger scenario (see also van~der~Wel et al.~\citeyear{vanderwel09}; Barro et al.~\citeyear{barro13}).

Secondly, we discuss the shapes of the galaxies by using $n$.
As the shape of galaxy is believed to exhibit the evolution trace, irrespective of the galaxy masses, we discuss the evolution of $n$.
In Fig.~\ref{fig:fig.10}, we see that the typical $n$ of QGs is larger than that of SFGs.
There is also difference between the evolution of $n$ of massive and less massive QGs, 
though the latter are limited to $z\leq1.5$ due to the completeness limit.
We found that $n$ of massive QGs significantly evolved with $\alpha_n\sim0.74$ of $n\propto (1+z)^{-\alpha_n}$, while those of less massive QGs and SFGs were unchanged on average ($\alpha_n\sim0.17$ and $\sim 0$, respectively).
The findings of the evolution in $n$ would give us a clue to understanding the morphological evolution of high-$z$ galaxies.
\textcolor{black}{
One of the most favorable scenarios is the inside-out mass growth, where the mass accretion in the outer parts of compact galaxies increase $r_\mathrm{e}$ and $n$ toward low-$z$ (van~Dokkum et al.~\citeyear{vandokkum10}).
In van~Dokkum et al., they showed the evolution of $n$ ($n\sim4$ to $\sim2$ from $z\sim2$ to $\sim0.5$), as well as the size evolution ($r_\mathrm{e}\sim3$~kpc to 8~kpc over the same redshift range). 
}
In addition to the observational results, numerical simulation by Naab et al.~(\citeyear{naab09}) explained the evolution of $n$ by minor merging of satellite galaxies.
As done in Chevance et al.~(\citeyear{chevance12}), the investigation for the distribution and evolution in of $b/a$ over the redshift range may help our understandings for the low-$n$ QGs, which could be the progenitor of local bulges (see also Trujillo, Carrasco \& Ferr\'e-Mateu~\citeyear{trujillo12}), though the small sample in the present study would not give the robust conclusion.

Based on the results of $r_\mathrm{e}$ and $n$ in the present study, we then discuss the formation and evolution of massive QGs, taking account of the evolution of less massive QGs and SFGs as well.
First, focusing on $n$, less massive SFGs at the whole redshift range ($0.5\leq z\leq2.5$) are thought to be the progenitor of the less massive QGs after exhausting their gas and passively evolve with little change in size and shape.
In the context, the progenitors of comparatively less massive QGs at high-$z$ could be small or amorphous SFGs located at higher redshift ($z\gtrsim3.0$) (Kajisawa \& Yamada~\citeyear{kajisawa01}), most of which are not included in the present study due to our detection limit.
\textcolor{black}{
While the sizes of QGs are smaller than those of SFGs at the whole redshift, it is believed that there are shrinkage of extended SFGs when they transform into QGs (Barro et al.~\citeyear{barro13b}).
The shrinkage of SFGs is also investigated by Dekel et al.~(\citeyear{dekel13}), in which migration of star-forming clumps forms the blue compact galaxies (or blue nuggets), then quench into red nuggets (see also Noguchi~\citeyear{noguchi99}).
We need more detailed investigation whether the compact galaxies keep $n$ lower after the migration of star-forming clumps, while Williams et al.~(\citeyear{williams14}) showed that blue nuggets had $\langle n\rangle \sim$ 2-3 and red nuggets $\langle n\rangle \sim$ 3-4.
Whitaker et al.~(\citeyear{whitaker12}) found very little difference in the sizes of young and old quenched galaxies.
}

After the birth of low-$n$ red QGs, in the later epoch ($z<1.5$) major or minor mergers of those dry (gas-poor) QGs dominates the evolution of massive QGs, enlarging their size rapidly over the cosmic time (Gao et al.~\citeyear{gao04}).
The dry merger is also consistent with the evolution of $n$ because it is efficiently change the light profile of galaxy (Barnes~\citeyear{barnes92}; Hernquist~\citeyear{hernquist92}; Naab et al.~\citeyear{naab09}), while wet (gas-rich) merger would reproduce disk-like galaxies (Steinmetz \& Navarro~\citeyear{steinmetz02}; Springel \& Hernquist~\citeyear{springel05}; Robertson et al.~\citeyear{robertson06}; Cox et al.~\citeyear{cox06}).
On the other hand, we found no significant evolution in $r_\mathrm{e}$ and $n$ of less massive QGs.
\textcolor{black}{
To explain this, we might need to consider the environmental effect on their formation (or halo mass size; Dekel et al.~\citeyear{dekel13}).
However, we avoid discussing this because of the small sample and limited redshift range of the less massive QGs in the present study.
}

\begin{figure}
\figurenum{12}
\plotone{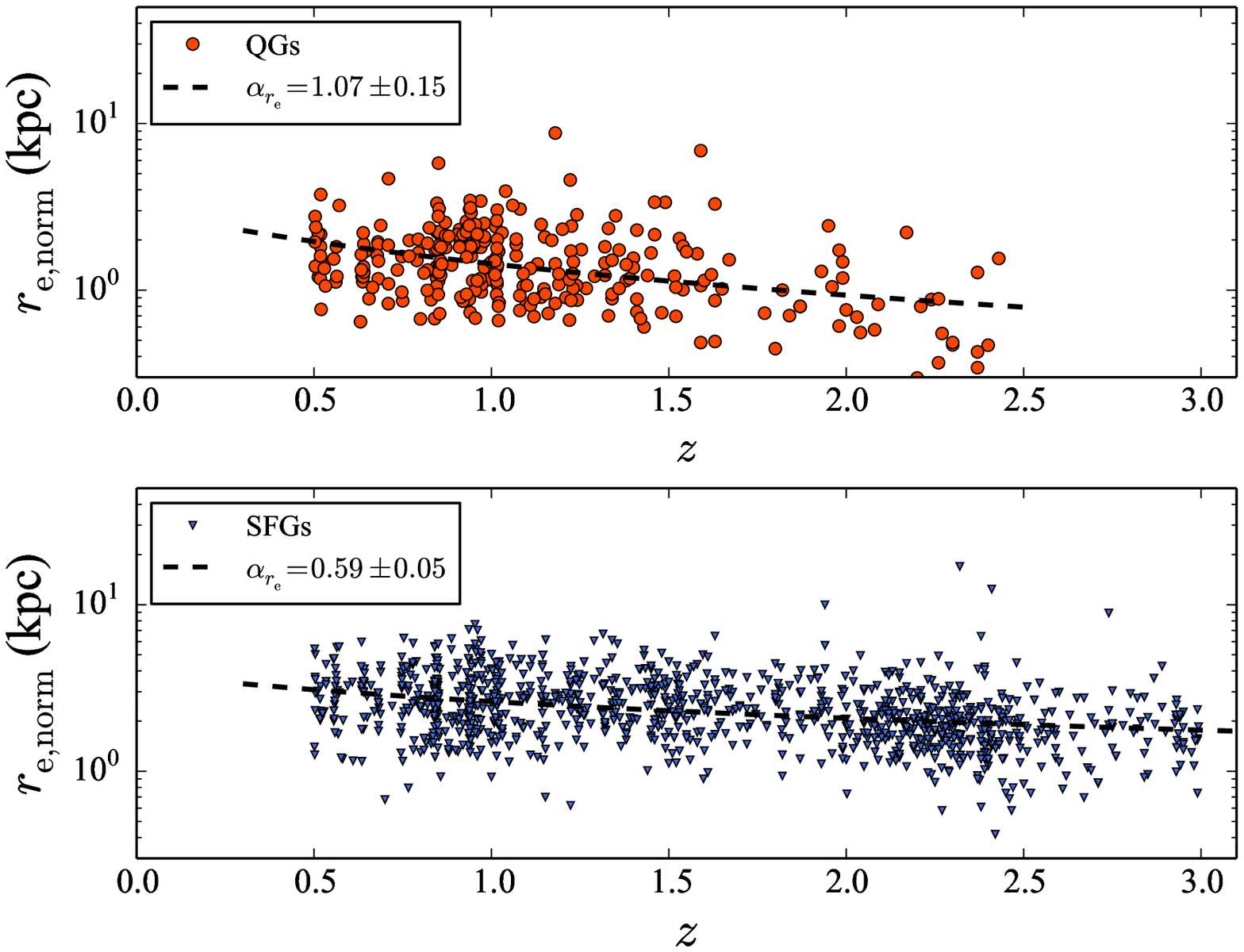}
\caption{
Mass-normalized size evolution for QGs (top) and SFGs (bottom).
Black dotted lines represent the regressions of $r_\mathrm{e,norm}\propto(1+z)^{-\alpha_{r\mathrm{e}}}$.
}
\label{fig:norm}
\end{figure}

\textcolor{black}{
It should be noted that the argument above (see also Carollo et al.~\citeyear{carollo13}) is complicated by the fact that samples of massive galaxies, not separated into star-forming and quiescent, have been demonstrated to grow smoothly with time (e.g., van~Dokkum et al.~\citeyear{vandokkum10}; Patel et al.~\citeyear{patel13}).
To corroborate the evolution scenario avobe, the discussion based on comoving number density for each group of galaxies would be helpful.
Patel et al.~(\citeyear{patel13}) studied the structural evolution of massive galaxies by linking progenitors and descendants at a constant cumulative number density.
However, small number of samples in the present study would not be enough for the robust discussion in statistical sense.
The field variance in a small field would also hamper the reliable conclusion.
In addition, the present study investigated the evolution of $r_\mathrm{e}$ at given two mass bins while those previous studies investigated the evolution taking account of the stellar mass evolution ({\it differential size evolution}) by using the constant number density method or mass-normalized size.
The mass-normalized size is useful not only to see the differential size evolution, but also to compensate the weak completeness, which could give rise to the biased (or spurious) size evolution.
In Fig.~\ref{fig:norm} we show the mass-normalized size evolution for QGs and SFGs.
$r_\mathrm{e,norm}$ is derived as $r_\mathrm{e,norm}=r_\mathrm{e}/(M_*^\mathrm{cor}/M_\mathrm{c})^{\alpha_\mathrm{M}}$, where ${\alpha_\mathrm{M}}$ is the linear slope of the size-stellar mass relation for each redshift bin (see Section~\ref{sec:sec5.1}).
It is noted that the relations are derived only using the sample within the completeness limits.
Then, we see the size evolution with $r_\mathrm{e}\propto(1+z)^{-\alpha_{r_\mathrm{e}}}$ finding that $\alpha_{r_\mathrm{e}}=1.07\pm0.15$ for QGs and $\alpha_{r_\mathrm{e}}=0.59\pm0.05$ for SFGs, which is consistent with the results of massive QGs and SFGs at fixed mass bins.
The results ensure our results of the size evolution over the redshift ranges, while we need more sample to investigate the weak evolution of $r_\mathrm{e}$ (and $n$) of less massive QGs.
}

%%%%%%%%
% Graphics %
%%%%%%%%

Finally, we should refer to the scatters in size at whole redshift range comparable to that of the local populations (Fig.~\ref{fig:fig.9}).
This suggests that there had already been massive QGs at $z\sim$ 1.5--3, whose morphological properties were similar to that of the local ones.
This is encouraged by the fact that the velocity dispersion of QGs at $z\sim1.6$ were found to be comparable to those of the local galaxies (Onodera et al.~\citeyear{onodera12}).
It should be noted, however, that a single S\'ersic profile may not be appropriate to fit galaxies at high-$z$, even if they seem to be well-virialized.
Bruce et al.~(\citeyear{bruce12}) applied {\ttfamily GALFIT} with two components (bulge, disk) massive galaxies ($M_*>10^{11}$~M$_{\sun}$) at $1<z<3$ and found many bulge$+$disk systems.
In the local universe, Huang et al.~(\citeyear{huang13}) found the elliptical galaxies were well fitted by three components.
However, multi-component fitting for our sample is too complicated because we of the small apparent size of, in particular, galaxies with $M_*\sim10^{10}$~M$_{\odot}$ at $z>1$, making it hard to find the "best-fit" results, while we adopted a single component S\'ersic fit in the present study.
\textcolor{black}{
To discuss the origin of the scatters in size, we need further investigation with, for example, mass-normalized size, which is beyond the present study.
}

%%%%%%%%%%%%

\acknowledgments
\textcolor{black}{
We would like to thank the referee for his/her helpful and constructive comments, which served to greatly improve the paper.
}
We would like to give special thanks to Chien Y. Peng for his useful advise about the morphological fitting with {\ttfamily GALFIT}.
We thank Pierre Marchand for his advise about {\ttfamily GALFIT} pipeline program codes, Taira Oogi and Kohei Hayashi for useful discussion, Arjen van~der~Wel for his advice about drizzled PSFs, and Jennifer Mack and Andy Fruchter for their advise about drizzle tasks.
This work has been supported in part by a Grant-in-Aid for Scientific Research (24253003) of the Ministry of Education, Culture, Sports, Science and Technology in Japan. 
This work is based on observations taken by the CANDELS Multi-Cycle Treasury Program with the NASA/ESA HST, which is operated by the Association of Universities for Research in Astronomy, Inc., under NASA contract NAS5-26555. 
MODS catalog has been accomplished by MOIRCS builders. 
We owe the present study to their dedicated efforts. 

{\it Facilities:} \facility{$HST$ (STIS)}, \facility{Subaru/MOIRCS ($NAOJ$)}.
%%%%%%%%%%

\appendix

\section{A. Difference between {$r_\mathrm{e}$} and {$r_{50}$}}\label{sec:app}
\begin{figure*}
\begin{center}
\figurenum{A1a}
\plotone{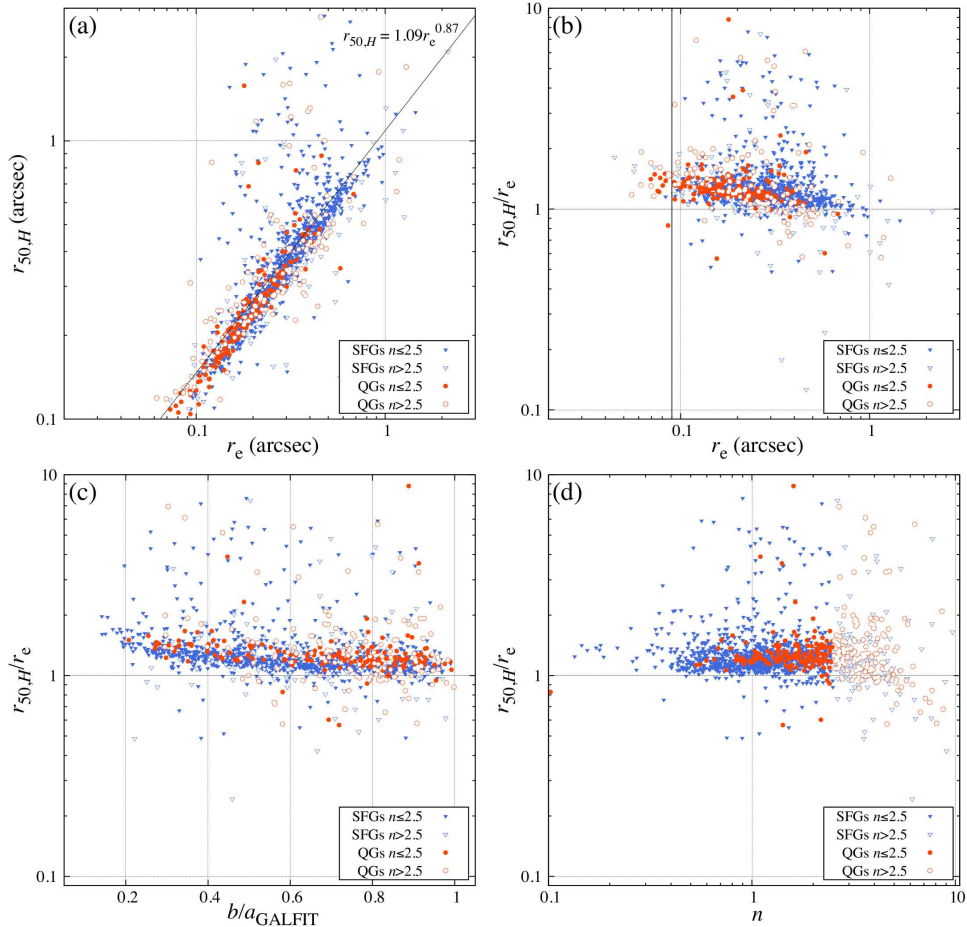}
\caption{\label{fig:fig12a}
Comparison of $r_\mathrm{e}$ and $r_{50,H}$ for the samples in the present study, where $r_{50,H}$ is $H_{160}$-band half-light radius obtained by {\ttfamily SExtractor}.
The symbols represents SFGs and QGs with $n\leq2.5$ (filled blue triangles and red circles, respectively) and those with $n>2.5$ (open blue triangles and red circles, respectively).
{\bf (a)} Comparison of $r_\mathrm{e}$ and $r_{50,H}$ with a regression of ${r_{50,H}}=a r_\mathrm{e}^b$ with $a$=1.09 and $b$=0.87.
{\bf (b)} Comparison of $r_{50,H}/r_\mathrm{e}$ as a function of $r_\mathrm{e}$.
The vertical solid line at $r_\mathrm{e}=0\arcsec\!.09$ is seeing radius for the $H_{160}$-band image.
{\bf (c),(d)} Same comparisons as (b) but as a function of $b/a$ and $n$, respectively.
}
\end{center}
\end{figure*}

In this section, we refer to the discrepant results based on non-parametric half-light radius.
Ic12 obtained non-parametric $r_{50,K}$ by {\ttfamily SExtractor} with the ground-based $K_\mathrm{s}$-band images to study the mass-size relation and claimed much weaker size evolution for massive galaxies in the present region, irrelevant to galaxy populations.
Since their results would be concerned for seeing effect ($\sim$ 0\arcsec\!.5), we obtained again the half-light radius ($r_{50,H}$) on WFC3 $H_{160}$ image with smaller PSF.
The PSF effect is corrected for with,
\begin{equation} \label{eq:eq7}
r_{\mathrm{50},H}=\sqrt{{r_{50,H,\mathrm{original}}}^2-{r_\mathrm{PSF}}^2},
\end{equation}
where $r_{50,H,\mathrm{original}}$ is {\ttfamily SExtractor}'s original output and $r_\mathrm{PSF}$ is the PSF radius, here set to 2.55~pixels for the $H_{160}$ image following the same process as in Ic12.
In Fig.~\ref{fig:fig12a}, we compare the results of $r_{\mathrm e}$ and $r_{50,H}$.
(Such a comparison is also reported in Fig.~4 of {\ttfamily GALFIT} Q\&A.)
As shown in panels (a) and (b), $r_{50,H}$ and $r_\mathrm{e}$ are biased at smaller radius, where PSF would significantly affect.
The dependence on other parameters, such as $b/a$ and $n$, are clearly seen in panel (c) and (d) of Fig.~\ref{fig:fig12a}.
The dependence on $b/a$ in (c) is reasonably understood by the definition of two radii; in {\ttfamily SExtractor}, $r_{50}$ is defined as the radius of circular aperture which encircles half $H_\mathrm{AUTO}$, while {\ttfamily GALFIT} calculates $r_\mathrm{e}=a_\mathrm{e}\sqrt{b/a}$.

%%%%%%%%
% Graphics %
%%%%%%%%

\begin{figure*}
\begin{center}
\figurenum{A1b}
\plotone{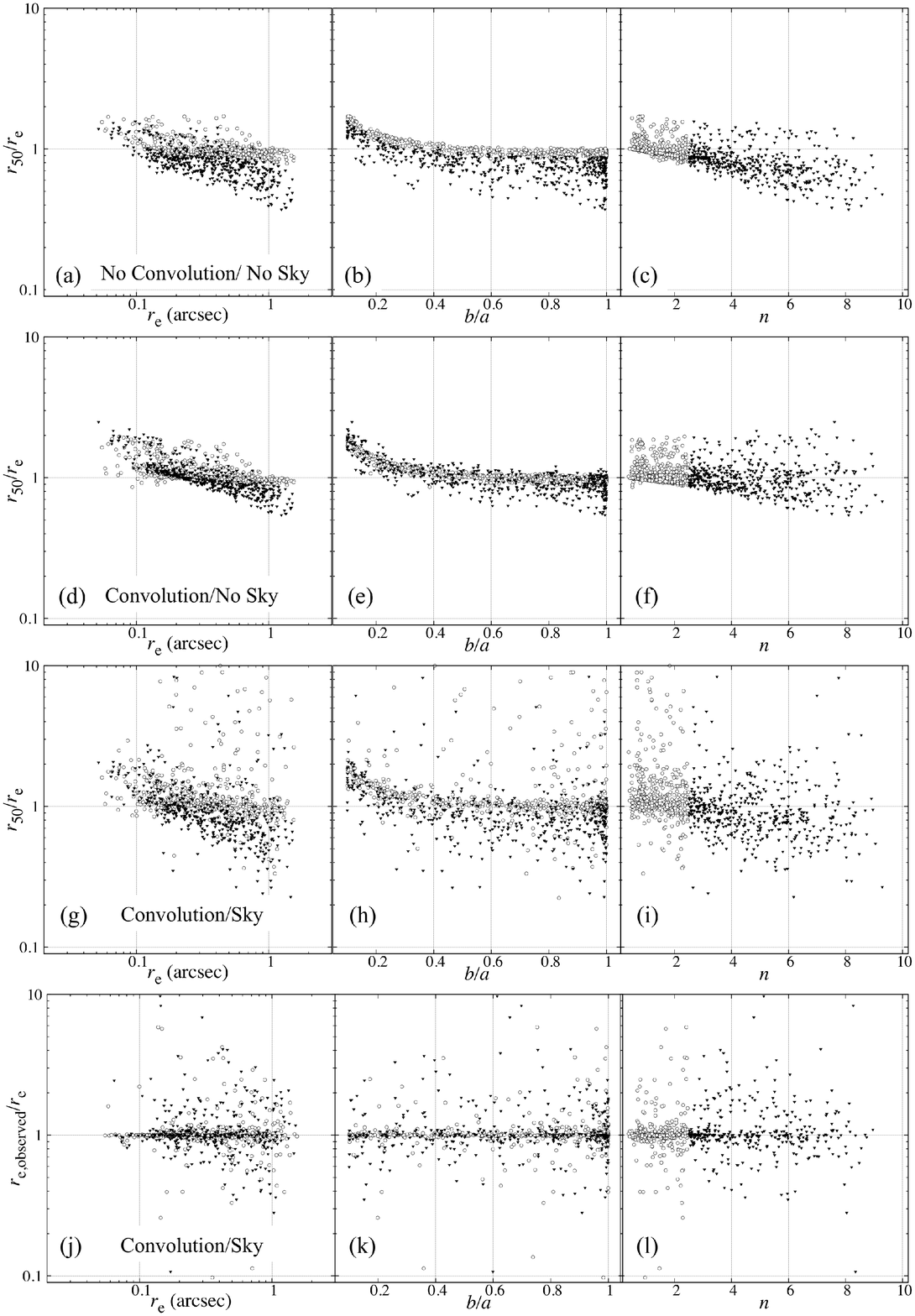}
\caption{\label{fig:fig12b}
Comparison of $r_\mathrm{e}$ and $r_{50}$ for three types of artificial galaxy, where $r_{50}$ is half-light radius obtained by {\ttfamily SExtractor}.
$r_\mathrm{e}$ used here is input value of the artificial galaxies.
The symbols represent AGs with $n<2.5$ (open circles) and $n\geq2.5$ (filled triangles).
$Top$~$panels$: Results for artificial galaxies without convolution and background sky noise.
Three horizontal panels show the comparison of $r_\mathrm{e}$ and $r_{50}$ as a function of $r_\mathrm{e}$, $b/a$ and $n$.
$Second$~$panels$: Results for artificial galaxies with convolution and without background sky noise.
$Third$~$panels$: Results for artificial galaxies with convolution and background sky noise.
$r_{50}$ for the convolved galaxies are corrected using Eq.~\ref{eq:eq7} in the main text.
$Bottom$~$panels$: Results of observed effective radius, $r_\mathrm{e,observed}$, instead of $r_{50}$, for the same galaxies as in third panels.
}
\end{center}
\end{figure*}

We further investigate the discrepancy between $r_{50}$ and $r_\mathrm{e}$ using the AGs in Section~\ref{sec:sec3.3}, of which results are shown in Fig.~\ref{fig:fig12b}.
We here prepare three types of AGs; without convolution and sky background noise, with convolution and without sky background noise, and, with convolution and sky background noise.
$r_{50}$ of convolved AGs is corrected with Eq.~\ref{eq:eq7}.
The bottom panels in Fig.~\ref{fig:fig12b} show the comparison of observed effective radius, $r_\mathrm{e, observed}$, where we find no strong dependence of $r_\mathrm{e,observed}/r_\mathrm{e}$ on the other parameters.
These tests suggest that the biased results come from not only PSF convolution but also from the other morphological parameters.

\begin{figure*}
\figurenum{A2}
\plotone{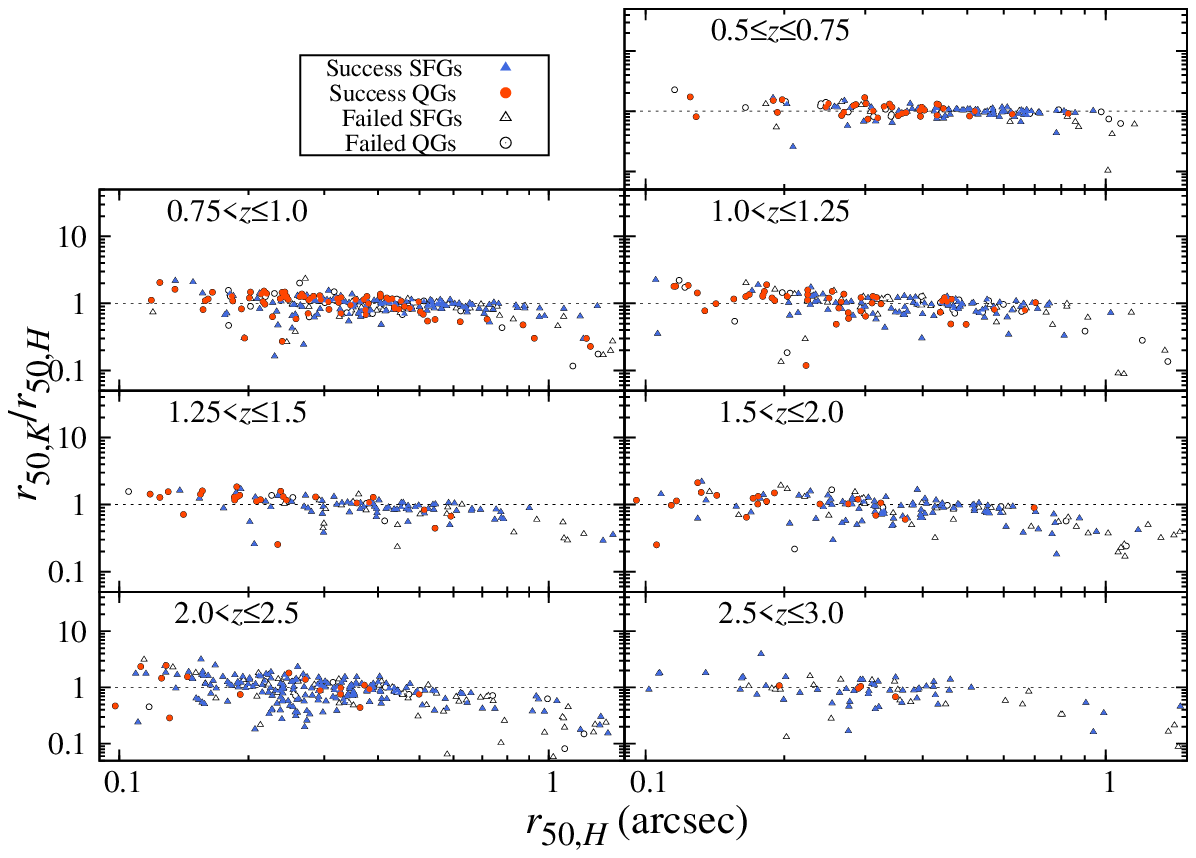}
\label{fig:fig.13}
\caption{
Comparison of $r_{50,H}$ and $r_{50,K}$ for the samples in the present study with $H_\mathrm{AUTO}\leq25$, where $r_{50,K}$ is $K_\mathrm{s}$-band half-light radius taken from Ic12.
Filled blue triangles and red circles represent successfully fitted SFGs and QGs, respectively.
Open triangles and circles represent unsuccessfully fitted SFGs and QGs, respectively.
}
\end{figure*}

To confirm the consistency in $r_{50}$ of the present study and Ic12, we compare $r_{50,H}$, which is corrected with Eq.~\ref{eq:eq7}, and $r_{50,K}$ of Ic12 in Fig.~\ref{fig:fig.13}.
We see no significant difference between them, irrespective of the PSF difference, except for the galaxies with large apparent radius which {\ttfamily GALFIT} failed to fit because of large chance to be contaminated by neighbors.
Figure~\ref{fig:fig.13} also includes the {\ttfamily GALFIT}-failed samples.

\begin{figure*}
\figurenum{A3}
\plottwo{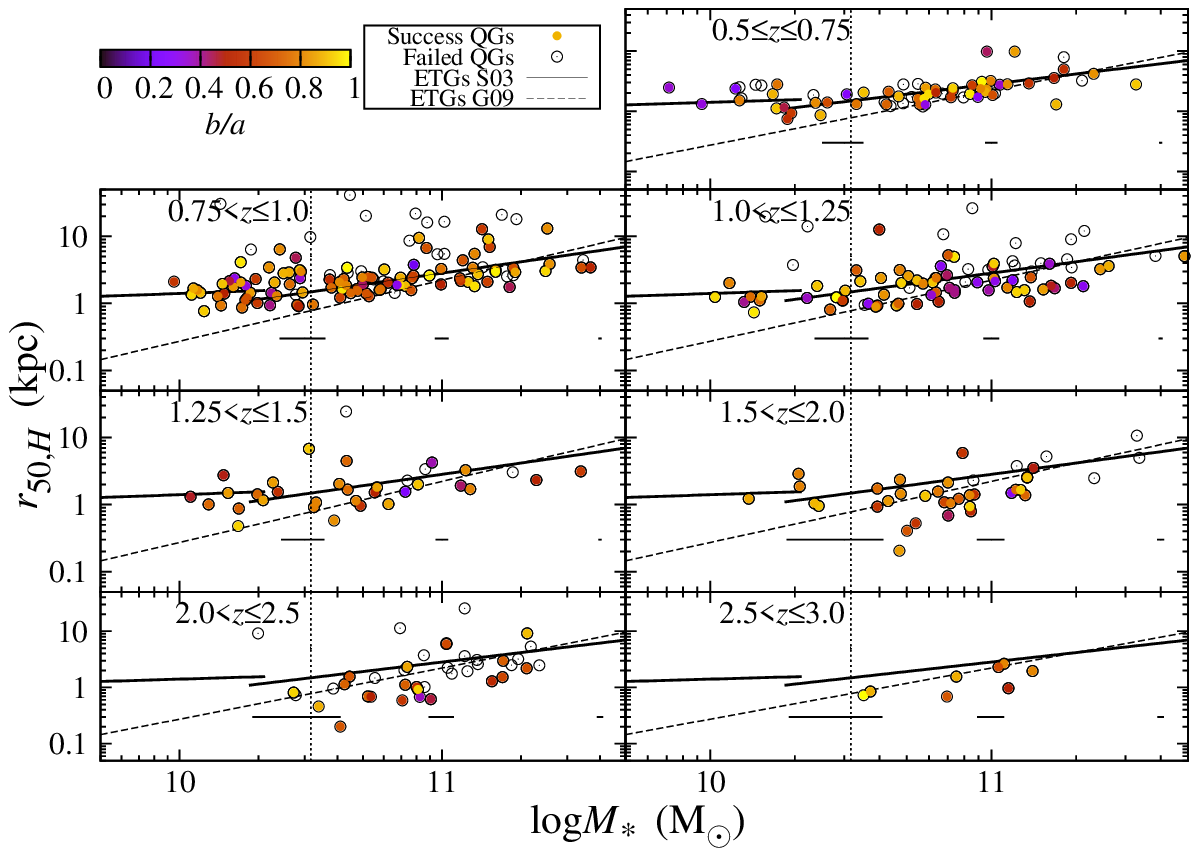}{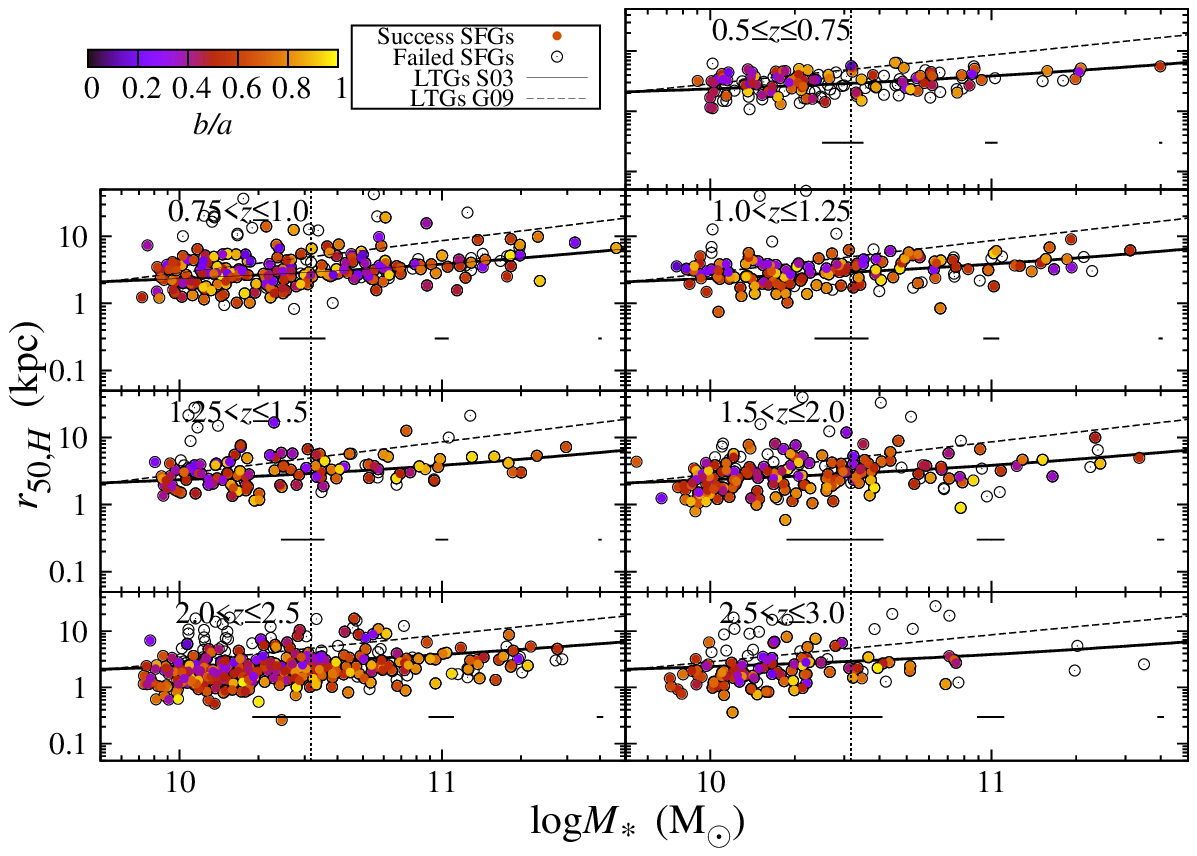}\label{fig:figa1a}
\caption{
{\it Left} : Stellar mass-size relations for half-light radius, $r_{50,H}$ of QGs obtained by {\ttfamily SExtractor}, including fitted (filled circles) and unfitted (open circles) samples by {\ttfamily GALFIT}.
{\it Right} : Same as the left panel, but for SFGs.
}
\end{figure*}

In Fig.~\ref{fig:figa1a}, we show the size-stellar mass relation in $r_{50,H}$.
Since $r_{50,H}$ was measured based on $H_\mathrm{AUTO}$, we use $M_*$ as the stellar mass rather than $M_*^\mathrm{cor}$.
In the figures, we see no significant difference among samples, which would deny the bias against compact galaxies, although $r_{50}$ would not represent $r_\mathrm{e}$ of {\ttfamily GALFIT}.
At a given mass, the size evolution is as mild as shown in Ic12 with $K_\mathrm{s}$-band image.
The size evolution with $r_\mathrm{e} \propto (1 + z)^{-\alpha_{r_\mathrm{e}}}$ for massive QGs $r_{50,H}$ is $\alpha_{r_\mathrm{e}}=0.64\pm0.19$, while that of the samples not including {\ttfamily GALFIT}-failed galaxies is $\alpha_{r_\mathrm{e}}=1.01\pm0.19$.
Visual inspection of {\ttfamily GALFIT}-failed galaxies in $H_{160}$ often shows neighbors or clumps, while those in ground-based $K_\mathrm{s}$-images are considered as a single object. 
The weak size-mass relations with non-parametric radius could be originated from those galaxies with neighbors or substructures.
However, it is noted, again, that we should take into account of the difference of size definitions between $r_\mathrm{e}$ and $r_{50}$.
{\ttfamily SExtractor} defines $r_{50}$ as an observed half-light radius of circular aperture, while {\ttfamily GALFIT} calculates $r_{\mathrm e}$ by assuming model galaxies with infinite size.
Calloro et al.~(\citeyear{carollo13}) stressed the possible biases in both
$r_\mathrm{e}$ and $r_{50}$ of galaxies due to, e.g., seeing effects and faint surface brightness.
As discussed above, the sizes of $r_{50}$ are found to be
consistent between grand-based $K_\mathrm{s}$ and WFC3 $H_{160}$
irrespective of the large PSF difference.
In the present study, the galaxies were chosen so as to have no
bias in $r_\mathrm{e}$.
Therefore, we also found no bias on the surface brightness.
$r_{50}$ of galaxies with larger $n$ would be prone to be
observed smaller than the real size
(e.g., Ichikawa et al.~\citeyear{ichikawa10}; Ic12; Calloro et al.~\citeyear{carollo13}).
If we correct the possible bias in $r_{50}$ to make them larger,
the size-mass relation for $r_{50}$ would lead to less size evolution,
which is more contradict to the significant size evolution in $r_\mathrm{e}$.
We see the evolution of both $n$ and $r_\mathrm{e}$ for QGs, though $r_\mathrm{e}$ could be biased if there is bias in $n$.
It should be noted that $H_\mathrm{AUTO}$ and $H_\mathrm{GALFIT}$ are consistent.
Probing the origin of the discrepancy would be beyond the
present scope.

%%%%%%%%
% Graphics %
%%%%%%%%

%%%%%%%%%%%%%%%%

\clearpage
%\LongTables % optionally
\begin{landscape}

\begin{deluxetable}{crrrrrrrr}

\setlength{\tabcolsep}{0.02in} 
\tabletypesize{\tiny}
\tablewidth{0pt}
\tablenum{2}
\tablecolumns{9}
\tablecaption{
Results of GALFIT simulation for AGs in $H_{160}$ image.\label{tb:tb2}
}
\tablehead{
\multicolumn{4}{c}{$n\leq2.5$} & \multicolumn{4}{c}{$n>2.5$}\\
\cline{2-9}
\colhead{$H_\mathrm{AUTO}$} & \colhead{$\Delta m$} & \colhead{$\Delta r_\mathrm{e}/r_\mathrm{e,input}$} & \colhead{$\Delta n/n_\mathrm{input}$} & \colhead{$\Delta (b/a)/(b/a_\mathrm{input})$} & \colhead{$\Delta m$} & \colhead{$\Delta r_\mathrm{e}/r_\mathrm{e,input}$} & \colhead{$\Delta n/n_\mathrm{input}$} & \colhead{$\Delta (b/a)/(b/a_\mathrm{input}$)}\\
\colhead{(mag)} & \colhead{$\times 10^{-2}$}& \colhead{$\times 10^{-2}$}& \colhead{$\times 10^{-2}$}& \colhead{$\times 10^{-2}$}& \colhead{$\times 10^{-2}$}& \colhead{$\times 10^{-2}$}& \colhead{$\times 10^{-2}$}& \colhead{$\times 10^{-2}$} }
\startdata
19--20	 &	-0.02	 $\pm$	0.54	 &	0.07	 $\pm$	0.79	 &	0.18	 $\pm$	1.08	 &	-0.02 	 $\pm$	0.42 	 &	-0.42	 $\pm$	1.06	 &	0.09	 $\pm$	1.61	 &	0.69	 $\pm$	1.81	 &	-0.15 	 $\pm$	0.56 	 \\
 20--21	 &	0.00	 $\pm$	0.63	 &	0.14	 $\pm$	0.58	 &	-0.10	 $\pm$	1.32	 &	-0.12 	 $\pm$	0.57 	 &	-0.54	 $\pm$	3.08	 &	1.16	 $\pm$	6.90	 &	0.97	 $\pm$	4.30	 &	-0.10 	 $\pm$	1.10 	 \\
 21--22	 &	-0.14	 $\pm$	2.42	 &	0.30	 $\pm$	2.61	 &	1.00	 $\pm$	3.26	 &	-0.28 	 $\pm$	0.89 	 &	-0.06	 $\pm$	5.51	 &	1.03	 $\pm$	10.29	 &	-0.19	 $\pm$	5.72	 &	-0.70 	 $\pm$	1.08 	 \\
 22--23	 &	-0.50	 $\pm$	5.00	 &	0.28	 $\pm$	4.81	 &	0.20	 $\pm$	7.56	 &	-0.87 	 $\pm$	4.29 	 &	1.34	 $\pm$	9.47	 &	-0.71	 $\pm$	14.15	 &	-3.08	 $\pm$	11.04	 &	-0.51 	 $\pm$	3.58 	 \\
 23--24	 &	0.74	 $\pm$	7.15	 &	-0.91	 $\pm$	6.64	 &	-1.44	 $\pm$	14.47	 &	-1.77 	 $\pm$	5.99 	 &	3.75	 $\pm$	16.18	 &	-6.18	 $\pm$	28.62	 &	-4.94	 $\pm$	15.91	 &	-1.01 	 $\pm$	7.58 	 \\
 24--25	 &	1.73	 $\pm$	16.1	 &	-1.99	 $\pm$	8.80	 &	-1.30	 $\pm$	25.99	 &	-1.88 	 $\pm$	11.60 	 &	5.41	 $\pm$	28.61	 &	-14.65	 $\pm$	34.04	 &	-10.02	 $\pm$	34.66	 &	-3.53 	 $\pm$	20.39 	 \\
 25--26	 &	8.98	 $\pm$	21.72	 &	-8.13	 $\pm$	22.1	 &	-18.81	 $\pm$	42.05	 &	-7.14 	 $\pm$	34.10 	 &	10.21	 $\pm$	50.00	 &	-19.02	 $\pm$	54.64	 &	-28.34	 $\pm$	55.15	 &	-17.38 	 $\pm$	15.06 	 \\
 26--27	 &	-7.35	 $\pm$	16.69	 &	0.89	 $\pm$	17.58	 &	26.97	 $\pm$	100.28	 &	-33.37 	 $\pm$	18.48 	 &	27.7	 $\pm$	43.22	 &	-35.63	 $\pm$	31.01	 &	-40.59	 $\pm$	36.26	 &	-33.57 	 $\pm$	28.04 	 \\
\enddata

\end{deluxetable}

\clearpage
\end{landscape}

\bibliography{adssample}
%\bibliographystyle{apj}
%\bibliography{ap-jour,reference}

\end{document}